\documentclass[11pt]{article}
\newcommand{\be}{\begin{equation}}
\newcommand{\ee}{\end{equation}}
\newcommand{\beq}{\begin{eqnarray}}
\newcommand{\eeq}{\end{eqnarray}}

\def\theequation{\arabic{section}.\arabic{equation}}
\usepackage{graphicx}
\begin{document}
%\pagestyle{empty}
%\begin{flushright}
%{BROWN-HET-1117} \\
%{March 1998}
%\end{flushright}
\begin{center}
{\bf\LARGE On the Summation of Feynman Graphs}\\
[7mm]
\vspace{1cm}  
H. M. FRIED
%\footnote{Supported in part by a Julian Schwinger Foundation Travel Grant }
\\
{\em Department of Physics \\
Brown University \\
Providence R.I. 02912 USA}\\
fried@het.brown.edu\\
tel:1-401-863-1467; fax: 1-401-863-3697\\
[5mm]
Y. GABELLINI%\footnote{Supported in part by a CNRS/Brown Accord }
\\
{\em Institut Non Lin\'eaire de Nice\\
UMR 6618 CNRS\\
 1361 Route des Lucioles\\
06560 Valbonne France}\\
yves.gabellini@inln.cnrs.fr\\
[5mm]

\vspace{4cm}
Abstract
\end{center}
A functional method to achieve the summation of all Feynman graphs relevant to a particular Field Theory process is suggested, and applied to  QED, demonstrating manifestly gauge invariant calculations of the dressed photon propagator in approximations of increasing complexity. These  lead in a natural way to the extraction of the leading logarithmic divergences of every perturbative order, and to a demonstration of the possible cancellation of all such divergences in the calculation of the (inverse of the) photon's wavefunction renormalization constant $Z_3$. This analysis provides a qualitative understanding of why the measured value of the renormalized fine structure constant is, approximately, 1/137.
\newpage
{\bf\section{Introduction}}
\setcounter{equation}{0}
Generations of Physicists have grown to maturity with the aid of Feynman graphs for various perturbative problems; and when the subjects of their theoretical researches have required a non--perturbative approach, they have attempted to sum sub--sets of Feynman graphs in order to build some semblance of a non--perturbative approach.

The only trouble with this approach is that it is physically and mentally impossible to sum all Feynman graphs needed for a particular amplitude. Consider, for example, a simple process of electron--electron scattering in QED for which one contemplates summing the contributions of all relevant Feynman graphs; and for simplicity let us neglect those ``~vacuum '' graphs with closed electron loops, as well as self--energy graphs for each electron, and consider only graphs built from the exchange of virtual photons between the two scattering electrons. As each order of perturbation theory is increased, a new class of graphs is generated, none of whose members can be obtained by iteration of the previous graphs; and as the perturbative order increases, there will be, eventually, an infinite number of graphs in each class. Bethe--Salpeter equations, whose kernel contains elements of a finite number of classes of graphs, are a so--called `` non--perturbative '' attempt at builing a solution which contains all powers of the coupling; but it cannot yield the correct answer because it has missed all the classes of graphs not contained in the kernel. And there is no known non--perturbative method of calculating that kernel.

Is there an analytic method of summing all relevant Feynman graphs ? The answer is positive, but it is a functional method, made possible by two distinct developments, both of which have been known for many years : the Schwinger--Symanzik \cite{one} functional expression for the Generating Functional (GF) of the Quantum Field Theory (QFT) in question; and the functional representations first given by Fradkin \cite{two} of the two functionals contained in that GF. The functional operations which yield all possible $n$--point functions of the theory are equivalent to a Gaussian--weighted functional integration over repeated powers of those two Fradkin functionals, depending on the process contemplated and on the degree of precision in which one is interested; an alternate and somewhat simpler `` linkage operator '' formalism can be employed which performs the same functions. In either case, the functional operation is defined over appropriate fluctuations of a space--time field $A(x)$, which are defined precisely by the Fradkin functional representations.

All of perturbation theory, which in any order is identically equivalent to the sum of all Feynman graphs of that order, may be obtained by a simple, perturbative expansion of the GF in powers of the relevant coupling constant; in any finite order of expansion of the GF, the functional operations may be performed exactly. But can the needed, non--perturbative functional operations be performed ? The answer is again positive, because the Fradkin representations are themselves Gaussian in their $A$ dependence, and an exact evaluation of a Gaussian weighted integration over Gaussian functional dependence can be performed. This effective summation over all Feynman graphs then produces an answer given in terms of the Fradkin functional representations of the two functionals alluded above; and one may well ask : Is this progress ?

Again, the answer is positive, and the reason is that the Fradkin representations are Potential Theory constructs, with decent approximations in different physical situations. In QED, for example, which QFT we shall here use to illustrate these remarks, the two functionals whose Fradkin representations are needed, are the electron Feynman (or causal) propagator $G_c(x,y| A)$ in a specified vector potential $A_{\mu}(x)$ :
$$\bigl[ m_0 +\gamma\!\cdot\!(\partial_x - ig_0A(x))\bigr]G_c(x,y|A) = \delta^{(4)}(x-y)$$
or, defining the operator $G_c[A]$ by : $<\!x\,| \,G_c[A] \,|\,y\!> \,= G_c(x,y|A)$ :
$$G_c[A] = \bigl[ m_0 +\gamma\!\cdot\!(\partial - ig_0A)\bigr]^{-1}$$
and the `` fermion determinant '', or `` vacuum functional '' $e\,^{\displaystyle L[A]}$, where : 
$$ L[A] = {\rm Tr}\ln\bigl[\, 1 - ig_0\gamma\!\cdot\!A\,S_c\,\bigr], \ \ \ S_c = G_c[0]$$
This fermion determinant represents the sum of all, single fermion loop graphs which contain all possible ( even ) numbers of attached photon lines.

The relativistic notation used throughout this article is the so--called ``east coast metric~'', with the scalar product defined by $a\!\cdot\!b = \vec a\!\cdot\!\vec b - a_0b_0$, and the Dirac matrices such that $\gamma_{\mu}^{\dagger} = \gamma_{\mu}$, $\gamma_{\mu}^2 = 1$ and $\{\,\gamma_{\mu}, \gamma_{\nu}\,\} = 2\delta_{\mu\nu}$.

\smallskip
At high energies, a very good approximation to $G_c[A]$ is a Bloch--Nordsieck, or eikonal approximation \cite{three}, and its various corrections, which are also ( and even simpler ) Gaussians. Corresponding approximations may be defined for $ L[A]$; but it is the special, exponentiated way in which $ L[A]$ appears which defines the nature of useful approximations; e.g., a functional cluster approximation.

These statements will here be illustrated by successive calculations of higher order corrections to the photon propagator of QED, ending with a simplified, intuitive extraction of the sum of the most relevant, divergent terms in the expression for the inverse of the photon's wave function renormalization constant. Gauge invariance in every order is manifest. In contrast to previous approximations which were unable to sum exact forms \cite{four}, we find a mechanism which has the possibility of producing a cancellation of the divergent perturbative logarithms, suggesting that $Z_3^{-1}$ may actually be finite, generating a finite charge renormalization; in the process, we find arguments as to why the renormalized fine structure constant is on the order of $1/137$, independently of the mass of the fermion ( electron or proton ) possessing that electric charge. The latter calculation is intuitive, rather than rigorous; but it is a compelling indication of the power of the functional approach described here.

To be more explicit, we briefly review in Section 2 the Schwinger--Symanzik functional expression for the Generating Functional (GF) of QED and the convenient version of the Fradkin representation for the functional fermion determinant. Then, in the context of this functional approach, we give a simple derivation of the lowest order vacuum polarization $\tilde K_{\mu\nu}^{(2)}(k)$. That calculation is then generalized in Section 3 to display the sum of all such $\tilde K_{\mu\nu}^{(2n)}(k)$, corresponding to the exchange of all possible virtual photons across and on the same fermion lines of that simplest closed fermion loop graph; and in the process we discuss two sets of special cancellations which simplify all subsequent calculations immensely. ( In contrast to the usual Feynman graph calculations, the cancellation of certain divergences is apparent immediately, before the relevant functional integral ( FI ) is evaluated ). We set up the old Jost--Luttinger \cite{five} calculation of $\tilde K_{\mu\nu}^{(4)}(k)$ in this new and simpler formalism, and leave it as an exercise for the interested reader.

Each of the $\tilde K_{\mu\nu}^{(2n)}(k)$ so calculated displays one or more log divergences, as does the photon's $Z_3^{-1}$. We introduce in Section 4 the Dominant Part ( DP ) Model, which extracts all possible log divergences from every order of the graphs defining 
\be \sum_{n=1}^\infty\tilde K_{\mu\nu}^{(2n)}(k)\ee
and indicate how and why the result is still divergent. At this stage, perturbative renormalization can be carried out in the conventional way.

But the complete $\tilde K_{\mu\nu}(k)$ contains an infinite number of closed fermion loops, with all their radiative corrections, attached in all possible ways to the graphs which gave $(1.1)$; and in Section 5 we introduce an Extended DP Model which extracts and sums the divergent parts of all graphs which contribute significantly to $Z_3^{-1}$. With the use of simplifying but reasonable approximations to the two resulting integrals, we then find that the $Z_3^{-1}$ so obtained is finite. The requirement that the Im$Z_3 = 0$ then defines an `` eigenvalue condition '' for $\alpha_0 = g_0^2/4\pi$; and together with the finiteness of $Z_3^{-1}$, the possible value(s) of $\alpha_0Z_3(\alpha_0)$ define the renormalized $\alpha$. Within the context of our seemingly reasonable approximations, it follows that $Z_3^{-1}(\alpha_0)$, and hence $(\alpha_0)$, are independant of the fermion mass; and that the expected value of $\alpha$ should be close to $1/137$.

Finally, in the Summary of Section 6, we compare this analysis with previous, unsuccessful ones of four decades ago. We emphasize that the arguments and proofs given in this paper are intuitive and heuristic; they are `` physicists proofs ''. We believe that they provide a framework in which one can resolve the long standing puzzle of whether QED, and other gauge theories, can be considered as finite and true theories of Nature.
\bigskip
{\bf\section{The Generating Functional and Fradkin's representation}}
\setcounter{equation}{0}
We begin by stating the Schwinger--Symanzik Generating Functional for QED, defined in a covariant gauge :
\beq\matrix{&\displaystyle{\cal Z}^{QED}[\eta, \bar\eta,J] = {1\over <\!0\,|S|\,0\!>}\,\exp\Bigl[ ig_0\!\int\!\!d^4x\,{\delta\over\delta \eta(x)}\biggl(\gamma^{\mu}{\delta\over\delta J^{\mu}(x)}\biggr){\delta\over\delta \bar\eta(x)}\Bigr]  \hfill\cr\noalign{\medskip} & \displaystyle \exp\Bigl[i\!\int\!\!d^4x\,d^4y\,\bar\eta(x)\,S_c(x-y)\,\eta(y) + { i\over2}\int\!\!d^4x\,d^4y\,J^{\mu}(x)\,D_{c,\mu\nu}(x-y)\, J^{\nu}(y)\Bigr] \hfill}\eeq
An alternate, equivalent, and somewhat more convenient representation is given by :
\beq\matrix{\displaystyle&{\cal Z}^{QED}[\eta, \bar\eta,J] =\displaystyle {1\over <\!0\,|S|\,0\!>}\,\exp\Bigl[ { i\over2}\int\!\!d^4x\,d^4y\,J^{\mu}(x)\,D_{c,\mu\nu}(x-y)\, J^{\nu}(y)\Bigr]\hfill\cr\noalign{\medskip} & \displaystyle\times e\,^{\displaystyle {\cal D}\!_A}\exp\Bigl[i\!\int\!\!d^4x\,d^4y\,\bar\eta(x)\, G_c(x,y|A)\,\eta(y)\Bigr]\exp\bigl[ L[A]\bigr]\,\hfill}\eeq
with $\displaystyle A_{\mu}(x) = \int\!\!d^4y\,D_{c,\mu\nu}(x-y)\,J^{\nu}(y)$. 

The normalization ${\cal Z}^{QED}[\,0, 0, 0\,] = 1$ is defined by : 
$<\!0\,|S|\,0\!>\,=\,e\,^{\displaystyle {\cal D}\!_A}\,\exp\bigl[ L[A]\bigr]\Bigl|_{A=0}$, and $e\,^{\displaystyle {\cal D}\!_A}$ represents the `` linkage operator '' :$$\displaystyle{\cal D}\!_A = -{i\over2}\,\int\!\!d^4x\,d^4y\,{\delta\over\delta A_{\mu}(x)}D_{c,\mu\nu}(x-y){\delta\over\delta A_{\nu}(y)}$$
The free photon propagator is given in the Feynman gauge by : $\displaystyle\tilde D_{c,\mu\nu}(k) = {\delta_{\mu\nu}\over k^2 - i\varepsilon}$ in momentum space, and $\displaystyle D_{c,\mu\nu}(z) = {i\over4\pi^2}\,{\delta_{\mu\nu}\over z^2 + i\varepsilon}$ in configuration space. 

\smallskip
In a previous paper dealing with both photon and electron propagators \cite{six}, use was made of the Fradkin functional representation for $G_c(x,y|A)$, in the calculation of the dressed electron propagator, but the calculation was performed in a `` quenched approximation '', negelecting effects of the vacuum functional $ L[A]$ :
$$ S_c'(x-y) = e\,^{\displaystyle{\cal D}\!\!_A}\,G_c(x,y\,| A)\,{e\,^{\displaystyle L[A]}\over < S >}\Bigl|_{A=0}\longrightarrow e\,^{\displaystyle{\cal D}\!\!_A}\,G_c(x,y\,| A)\Bigl|_{A=0}$$
In this paper, we treat radiative corrections to the fully dressed photon propagator :
$$\displaystyle D_{c,\mu\nu}'(x-y) = D_{c,\mu\nu}(x-y) + \int\!\!\!\int\!D_{c,\mu\lambda}(x-u)K_{\lambda\sigma}(u-w)D_{c,\sigma\nu}(w-y)du\,dw$$
\vskip-0.5cm
\beq{}&\displaystyle iK_{\mu\nu}(x-y)=e\,^{\displaystyle{\cal D}\!\!_A}\,{\delta\over\delta A_\mu(x)}{\delta\over\delta A_\nu(y)}\,e\,^{\displaystyle L[A]} /\!< S >\Bigl|_{A=0}\nonumber \\& =\displaystyle e\,^{\displaystyle{\cal D}\!\!_A}\,\biggl[{\delta^2 L\over{\delta A_\mu(x)\delta A_\nu(y)}} + {\delta L\over\delta A_\mu(x)}{\delta L\over\delta A_\nu(y)}\biggr]\,e\,^{\displaystyle L[A]} /\!< S >\Bigl|_{A=0}\eeq
in contrast to the quenched approximation of  ref.\cite{six} :
\beq \displaystyle iK_{\mu\nu}(x-y)=e\,^{\displaystyle{\cal D}\!\!_A}\,{\delta^2 L\over{\delta A_\mu(x)\delta A_\nu(y)}}\Bigl|_{A=0}\eeq
and find that our conclusions are strongly dependent upon the effects of $\exp\bigl[ L[A]\bigr]$.

We here sketch the evaluation of the lowest order $\tilde K_{\mu\nu}^{(2)}(k)$ directly from the gauge invariant $L[A]$ formalism, obtained from $(2.4)$ without the linkage operations of that equation. The fermion determinant has an exact Fradkin representation of the form :
\beq{}&\displaystyle L[A] = -{1\over 2}\int_0^{\infty}\!\!{ds\over s}\,e\,^{\displaystyle -ism_0^2}\,e\,^{\displaystyle i\int_0^s\!\!ds'\,{\delta^2\over\delta v_{\mu}^2(s')}}\,\delta^{(4)}\Bigl( \int_0^s\!ds'\,v(s') \Bigr)\int\!\!d^4x'\nonumber \\& \times\displaystyle \,e\,^{\displaystyle -ig_0\int_0^s\!\!ds'\,v_{\mu}(s')\,A_{\mu}( x' - \int_0^{s'}\!v)}\,{\rm tr}\biggl( e\,^{\displaystyle g_0\int_0^s\!\!ds'\,\sigma_{\mu\nu}\,F_{\mu\nu}( x' - \int_0^{s'}\!v)}\biggr)_+\Bigl|_{v=0} - \,L_0\eeq
where $F_{\mu\nu} = \partial_{\mu}A_{\nu} - \partial_{\nu}A_{\mu}$, $\sigma_{\mu\nu}$ denotes the gamma matrix combination ${1\over 4}[\gamma_{\mu}, \gamma_{\nu}]$, $v_{\mu}(s')$ denotes the four velocity of the fermion with an instantaneous proper time $s'$ and the constant $L_0$ is such that $L[0] = 0$. A more convenient version of this representation is obtained from the variable change $u_{\mu}(s') = \int_0^{s'}ds''v_{\mu}(s'')$, which yields th{e normalized functional integral :
\beq{}&\displaystyle L[A] = -{1\over 2}\int_0^{\infty}\!\!{ds\over s}\,e\,^{\displaystyle -ism_0^2}\,\int\!\!d^4x'\,N\!\int\!d[u]\,e\,^{\displaystyle {i\over2}\int u(2h)^{-1}u}\,\delta^{(4)}\Bigl( u(s) \Bigr)\nonumber \\& \times\,e\,^{\displaystyle -ig_0\int_0^s\!\!ds'\,u_{\mu}'(s')\,A_{\mu}( x' - u(s'))}\,{\rm tr}\,\biggl( e\,^{\displaystyle g_0\int_0^s\!\!ds'\,\sigma_{\mu\nu}\,F_{\mu\nu}( x' - u(s'))}\biggr)_+ - L_0\eeq
where $N^{-1} = \displaystyle\int\!d[u]\,e\,^{\displaystyle{i\over2}\int u(2h)^{-1}u}$, $<\!s_1\,|\,h^{-1}\,|\,s_2\!>\, = \displaystyle{\overrightarrow\partial\over\partial s_1}\,\delta(s_1-s_2){\overleftarrow\partial\over\partial s_2}$ and $h(s_1,s_2) = \inf(s_1,s_2) = {1\over2}( s_1 + s_2 - | s_1 -s_2 | ) = s_1\,\theta(s_2-s_1) + s_2\,\theta(s_1-s_2)$.

It will be most relevant to remind the reader of certain rigorous properties of $L[A]$, which quantity in fact depends only upon $F_{\mu\nu}$, and can be easily be written as such . This property immediately carries the consequence that currents induced in the vacuum,\break $<j_{\mu}(x)>\,=ig\,\delta L/\delta A_\mu(x)$, are to satisfy charge conservation : $\partial_{\mu}\!\!<j_{\mu}(x)>\, = 0$. In terms of the $K_{\mu\nu}$ of $(2.3)$ or $(2.4)$, this means that $\partial_{\mu}K_{\mu\nu} = \partial_{\nu}K_{\mu\nu} = 0$, so that the $\tilde K_{\mu\nu}(k)$ are expected to have the gauge invariant form $(k_{\mu}k_{\nu} - k^2\delta_{\mu\nu})\Pi(k^2)$. The simplest, order $\alpha = g^2/4\pi$, Feynman graph corresponding to a single closed fermion loop does not display this property; and in the past, special, ad hoc maneuvers were invented to restore gauge invariance. In the Fradkin representation for $L[A]$ used in this paper, gauge invariance to all orders is automatically satisfied.

It is important to note that there are two restrictions on the $u(s')$ variables, the first an implicit condition $u(0)=0$, which arises from the definition of $u(s')$; and the second condition, $u(s)=0$, explicitly stated by the delta function of $(2.6)$. For this lowest-order calculation, we replace $m_0$ by $m$.

Performing the pair of functional derivatives on $L[A]$ yields :
\beq{}&\displaystyle K_{\mu\nu}^{(2)}(x-y) = {i\over 2}\int_0^{\infty}\!\!{ds\over s}\,e\,^{\displaystyle -ism^2}\,N\!\int\!d[u]\,e\,^{\displaystyle {i\over2}\int u(2h)^{-1}u}\,\delta^{(4)}\Bigl( u(s) \Bigr)\nonumber \\& \times\displaystyle\,4g_0^2\int_0^s\!\!ds_1\int_0^s\!\!ds_2\,\biggl[- u_{\mu}'(s_1)u_{\nu}'(s_2) - (\partial_{\mu}\partial_{\nu} - \delta_{\mu\nu}\partial^2)\biggr]\delta^{(4)}(x-y+u(s_1) - u(s_2))\eeq
after making use of the trace properties tr$(1) = 4$, and tr$(\sigma_{\alpha\mu}\sigma_{\beta\nu}) = \delta_{\alpha\nu}\delta_{\beta\mu} - \delta_{\alpha\beta}\delta_{\mu\nu}$. Writing a Fourier representation of $\delta^{(4)}(x-y+u(s_1) - u(s_2))$, the second line of $(2.7)$ becomes :
\beq{}&\displaystyle {4g_0^2\over(2\pi)^4}\int\!\!d^4k\int_0^s\!\!ds_1\int_0^s\!\!ds_2\,e\,^{\displaystyle ik.(x-y) + ik.(u(s_1) - u(s_2))}\nonumber \\& \times\displaystyle\,\biggl[-u_{\mu}'(s_1)u_{\nu}'(s_2) + (k_{\mu}k_{\nu} - \delta_{\mu\nu}k^2)\biggr]\eeq
and one can see that gauge invariance is maintained by imagining the result of calculating $\partial_{\mu}$, or equivalently, of multiplying $(2.8)$ by $k_{\mu}$ and summing over $\mu$. The term coming from the sigma matrices obviously vanishes; and the $u'(s_1)$ dependence yields :

$$\displaystyle \int_0^s\!\!ds_1\, k.u'(s_1)\,e\,^{\displaystyle ik.u(s_1)} = -i\int_0^s\!\!ds_1\, {\partial\over\partial s_1}\,e\,^{\displaystyle ik.u(s_1)} = 0$$
because $u(0)=u(s)=0$. At this early stage, it is then clear that the result of this computation must be gauge invariant. 

In order to avoid any confusion on taking the $s_{1,2}$ derivatives of the $u'$ factors with the $u(s_1) - u(s_2)$ exponential terms, it will be useful to introduce the sources $g_{\mu}(s') = k_{\mu}[\delta(s'-s_1) - \delta(s'-s_2)]$, and also $f_{\mu}(s') = p_{\mu}\,\delta(s'-s)$, and there then follows :
\beq{}&\displaystyle \tilde K_{\mu\nu}^{(2)}(k) = 2ig_0^2\int_0^{\infty}\!\!{ds\over s}\,e\,^{\displaystyle -ism^2}\,\int_0^s\!\!ds_1\int_0^s\!\!ds_2\,\int\!\!{d^4p\over(2\pi)^4}N\!\int\!d[u]\,e\,^{\displaystyle {i\over2}\int u(2h)^{-1}u}\nonumber \\& \times\displaystyle \biggl[{\partial\over\partial s_a}{\partial\over\partial s_b}{\delta\over\delta g_{\mu}(s_a)}{\delta\over\delta g_{\nu}(s_b)} + (k_{\mu}k_{\nu} - \delta_{\mu\nu}k^2)\biggr]\,e\,^{\displaystyle i\int_0^s\!\!u.(f+g)}\Bigl|_{s_{a,b}\rightarrow s_{1,2}} \eeq
The normalized FI over the $u$ dependence is then immediate, and yields the exponential factor :
\beq e\,^{\displaystyle -i\int\!\!\!\int_0^s\!\!(f+g).h.(f+g)} \eeq
so that the combinations $\displaystyle{\partial\over\partial s_a}{\partial\over\partial s_b}{\delta\over\delta g_{\mu}(s_a)}{\delta\over\delta g_{\nu}(s_b)}$ generate :
\beq{}& \displaystyle  - 4\bigl[p_{\mu} + k_{\mu}(\theta(s_1-s_a)-\theta(s_2-s_a))\bigr]\bigl[p_{\nu} + k_{\nu}(\theta(s_1-s_b)-\theta(s_2-s_b))\bigr]\nonumber \\& \displaystyle \hskip-3truecm-2i\,\delta_{\mu\nu}\delta(s_a-s_b)\eeq
and in the limit as $s_a\rightarrow s_1$, $s_b\rightarrow s_2$, produce :
\beq \displaystyle (-2i)\delta_{\mu\nu}\delta(s_1-s_2) - 4\bigl[p_{\mu}p_{\nu} +{1\over4}k_{\mu}k_{\nu} + (p_{\mu}k_{\nu}+p_{\nu}k_{\mu})(\theta(s_1-s_2)-{1\over2})\bigr]\eeq
With the exponential factor of $(2.10)$ evaluated as :
\beq -isp^2-2is_{12}p.k-ik^2|s_{12}|\ ,\ \ \ \ \ s_{12} = s_1-s_2\eeq
the $\int\!d^4p$ may be performed, leading to the cancellation of the term coming from the tr$[\sigma_{\alpha\mu}\sigma_{\beta\nu}]$ dependence, and the result :
\beq{}&\displaystyle \tilde K_{\mu\nu}^{(2)}(k) = {g_0^2\over8\pi^2}\int_0^{\infty}\!\!{ds\over s}\,e\,^{\displaystyle -ism^2}\,{1\over s^2}\int_0^s\!\!ds_1\int_0^s\!\!ds_2\nonumber \\& \times\displaystyle \biggl[2i\,\delta_{\mu\nu}{1\over s} - 2i\,\delta_{\mu\nu}\,\delta(s_1-s_2) + 4\,{|s_1-s_2|\over s}\,(1- {|s_1-s_2|\over s})\,k_{\mu}k_{\nu} - \delta_{\mu\nu}k^2\biggr]\nonumber \\&  \displaystyle\times\exp\Bigl[-isk^2{|s_1-s_2|\over s}(1- {|s_1-s_2|\over s})\Bigr]\eeq
Finally, with the aid of the relations :
$$\int_0^s\!\!ds_1\int_0^s\!\!ds_2\,f(|s_1-s_2|) = 2s^2\int_0^1\!\!dy\,(1-y)\,f(sy)\ \ \ \ {\rm with}\ \ y = {|s_1-s_2|\over s}$$
$$\displaystyle\int_0^1\!\!dy\,(1-2y)\,e\,^{\displaystyle -isk^2y(1-y)}={i\over sk^2}\int_0^1\!\!dy\,{\partial\over\partial y}\,e\,^{\displaystyle -isk^2y(1-y)}=0$$
and :
$$\displaystyle\int_0^1\!\!dy\,y(1-2y)\,e\,^{\displaystyle -isk^2y(1-y)}= {i\over sk^2}\Bigl[1 - \int_0^1\!\!dy\,e\,^{\displaystyle -isk^2y(1-y)}\Bigr]$$
one obtains the result :
\beq \displaystyle \tilde K_{\mu\nu}^{(2)}(k) = (k_{\mu}k_{\nu} - \delta_{\mu\nu}k^2)\Pi^{(2)}(k^2)\ ,\ \ \ \ \Pi^{(2)}(k^2) = \int_0^{\infty}\!\!{ds\over s}\,e\,^{\displaystyle -ism^2}\,\Pi^{(2)}(k^2,s)\eeq
in which the gauge symmetry has been preserved, and the subsequent $\int\!ds$ leads to a log divergence at its lower limit. With :
$$\displaystyle\Pi^{(2)}(k^2,s) = {g_0^2\over 2\pi^2}\int_0^1\!\!dy\,y(1-y)\,e\,^{\displaystyle -isk^2y(1-y)}$$
one has the simplest vacuum polarization result of a half--century ago \cite{seven}. Then, with the inverse of the photon's wave function renormalization constant given by :
$$Z_3^{-1} = 1 + \Pi^{(2)}(0)$$
the renormalized, to order $\alpha$, vacuum polarization is given by :
$$\Pi^{(2)}(k^2) - \Pi^{(2)}(0)$$
which may easily be transformed into the more familiar form :
\beq\displaystyle\Pi^{(2)}(k^2) - \Pi^{(2)}(0) = -{2\alpha\over \pi}\int_0^1\!\!dy\,y(1-y)\,\ln\Bigl(\displaystyle 1 + y(1-y){k^2\over m^2}\Bigr)\eeq
The real beauty of this calculation of this old and familiar result is that the UV divergence does not appear until the very last step, the integration over the proper time; and so does not disrupt the gauge symmetry of its elements. In contrast, the familiar Feynman graph computation in momentum space is so badly divergent that the underlying gauge symmetry is lost, and must be reinstated by other means. This was, of course, known to Schwinger, who originated proper time calculations in QFT; but it is made clear upon employing the elegant and most useful representations of Fradkin.
\bigskip
{\bf\section{Radiative Corrections}}
\setcounter{equation}{0}
In this Section, we display the radiative corrections to the simplest, one closed fermion loop, and describe the cancellations which appear even before the corresponding FI is evaluated. These radiative corrections correspond to the action of the linkage operator acting upon the omitted $A$ dependence of Section 2, and can be succintly written by inserting the terms :
\beq e\,^{\displaystyle{\cal D}\!\!_A}\,e\,^{\displaystyle -ig_0\int_0^s\!\!ds'\,u_{\mu}'(s')\,A_{\mu}( x' - u(s'))}\,{\rm tr}\,\biggl( e\,^{\displaystyle g_0\int_0^s\!\!ds'\,\sigma_{\mu\nu}\,F_{\mu\nu}( x' - u(s'))}\biggr)_+\Big\vert_{A\rightarrow 0}\eeq
under all of the integrals of $(2.6)$. Eqs $(2.7)$--$(2.9)$ are still relevant, but before the $\int\!d^4p$ and the $\int\!d[u]$ are performed, the $u$ dependence of $(3.1)$ must be extracted. For this, $(3.1)$ may be rewritten, in the convenient form :
\beq \Biggl(e\,^{\displaystyle{\cal D}\!\!_A}\,e\,^{\displaystyle -ig_0\int_0^s\!\!ds'\,u'.A}\Biggr)\,e\,^{\displaystyle\buildrel\leftrightarrow\over{\cal D}\!\!_A}\,\Biggl(e\,^{\displaystyle{\cal D}\!\!_A}\,{\rm tr}\,\Bigl(e\,^{\displaystyle g_0\int_0^s\!\!ds'\,\sigma.F}\Bigr)_+\Biggr)\Big\vert_{A\rightarrow 0}\eeq
where $\displaystyle\buildrel\leftrightarrow\over{\cal D}\!\!_A = -i\!\int\! {\overleftarrow\delta\over\delta A}D_c{\overrightarrow\delta\over\delta A}\cdot$

The first simplification to be noted was proven in Appendix B of ref.\cite{six}, $$e\,^{\displaystyle{\cal D}\!\!_A}\,{\rm tr}\,\Bigl(e\,^{\displaystyle g_0\int_0^s\!\!ds'\,\sigma.F}\Bigr)_+\Big\vert_{A\rightarrow 0} = {\rm tr}\,1 = 4$$
and can be trivially generalized to the more relevant statement :
$$e\,^{\displaystyle{\cal D}\!\!_A}\,{\rm tr}\,\Bigl(e\,^{\displaystyle g_0\int_0^s\!\!ds'\,\sigma.F}\Bigr)_+ = {\rm tr}\,\Bigl(e\,^{\displaystyle g_0\int_0^s\!\!ds'\,\sigma.F}\Bigr)_+$$
so that the self linkages acting on this OE factor exactly cancel, to all orders in the coupling. In Feynman graph language, this would correspond to momentum space cancellations occurring in every higher order; in the Fradkin representation, one sees them immediately.

The self linkages $e\,^{\displaystyle{\cal D}\!\!_A}\,e\,^{\displaystyle -ig_0\int_0^s\!\!ds'\,u'.A}$ produce the dependence :
\beq \displaystyle e\,^{\displaystyle i{g_0^2\over2}\int_0^s\!\!ds_1\int_0^s\!\!ds_2\,u_{\mu}'(s_1)D_{c,\mu\nu}(u(s_1) - u(s_2))u_{\nu}'(s_2)}\,e\,^{\displaystyle -ig_0\int_0^s\!\!ds'\,u'.A} \eeq
leaving the cross linkage operation :
\beq\displaystyle e\,^{\displaystyle -ig_0\int_0^s\!\!ds'\,u'.A}\,e\,^{\displaystyle\buildrel\leftrightarrow\over{\cal D}\!\!_A}\,{\rm tr}\,\Bigl(e\,^{\displaystyle g_0\int_0^s\!\!ds'\,\sigma.F}\Bigr)_+\Big\vert_{A\rightarrow 0} \eeq
to be evaluated. The $A$ dependence inside the OE can be extracted by writing the latter as~:
$$\Bigl(e\,^{\displaystyle 2g_0\int_0^s\!\!\partial_{\mu}\,A_{\nu}\,\sigma_{\mu\nu}}\Bigr)_+=e\,^{\displaystyle -2ig_0\int_0^s\!\!ds'\,\partial_{\mu}\,A_{\nu}(x'-u(s')){\delta\over\delta\chi_{\mu\nu}(s')}}\,\Bigl(e\,^{\displaystyle i\int_0^s\!\!ds''\,\sigma_{\mu\nu}\chi_{\mu\nu}(s'')}\Bigr)_+$$
where $\chi_{\mu\nu}$ and $\delta/\delta\chi_{\mu\nu}$ are antisymmetric in $\mu$ and $\nu$. In this way, the cross linkages produce the term :
\beq \displaystyle e\,^{\displaystyle 2ig_0^2\int_0^s\!\!ds_1\int_0^s\!\!ds_2\,u_{\mu}'(s_1)\,\partial_{\lambda}\,D_{c,\mu\nu}(u(s_1) - u(s_2)){\delta\over\delta\chi_{\lambda\nu}(s_2)}}\eeq
which is then to act upon the waiting $\Bigl(e\,^{\displaystyle i\int_0^s\!\!ds'\,\sigma_{\mu\nu}\chi_{\mu\nu}(s')}\Bigr)_+$ factor. We recall that, in the Feynman gauge,  $D_{c,\mu\nu}(u(s_1) - u(s_2)) =\displaystyle{i\over4\pi^2}\,\delta_{\mu\nu}\,[(\Delta u)^2+i\varepsilon]^{-1}$ and $\Delta u_{\mu} = u_{\mu}(s_1) - u_{\mu}(s_2)$.

Repeating the argument given following $(2.8)$, one can immediately see that the terms of both $(3.3)$ and $(3.5)$ are individually and manifestly gauge invariant. We would like to argue that the contributions of the exponential of $(3.5)$ vanishes upon the evaluation of the FI, using reasoning very much like that of the Conjecture of ref.\cite{six}, wherein a rescaling of the variable $u_{\mu}(s')\rightarrow\sqrt Q\,\bar u_{\mu}(s')$, $Q$ real and greater than 1, produces a factor $1/Q$ multiplying the term of that exponential. In \cite{six}, where $u(s) + z = 0$, a related scaling operation, as $Q$ was increased, led to the effective replacement of the term corresponding to $(3.5)$ by a gauge dependent exponential factor proportional to $\ln Q$ appearing outside of its FI. In the present gauge invariant case, where $z=0$, such a rescaling changes $(3.5)$ to :
\beq \displaystyle e\,^{\displaystyle {g_0^2\over\pi^2Q}\int\!\!\!\int_0^s\!\!ds_1ds_2\,\bar u_{\mu}'(s_1)\,{\Delta\bar u_{\lambda}\over[(\Delta\bar u)^2+i\varepsilon]^2}{\delta\over\delta\chi_{\lambda\mu}(s_2)}}\,{\rm tr}\,\Bigl(e\,^{\displaystyle i\int\!\!\sigma.\chi}\Bigr)_+\Big\vert_{\chi\rightarrow 0}\eeq
or to :
\beq \displaystyle e\,^{\displaystyle {g_0^2\over\pi^2}\int\!\!\!\int_0^s\!\!ds_1ds_2\,\bar u_{\mu}'(s_1)\,{\Delta\bar u_{\lambda}\over[(\Delta\bar u)^2+i\varepsilon]^2}{\delta\over\delta\bar\chi_{\lambda\mu}(s_2)}}\,{\rm tr}\,\Bigl(e\,^{\displaystyle i/Q\int\!\!\sigma.\bar\chi}\Bigr)_+\Big\vert_{\bar\chi\rightarrow 0}\eeq
As $Q$ becomes arbitrarily large, these exponential terms become arbitrarily small, while the $Q$ dependence buried in the remainder of the FI is such that the result of the FI, with $(3.6)$ or $(3.7)$ replaced by unity, is independent of $Q$, regardless of the magnitude of $Q$. In other words, such an interchange of limits, taking $Q$ arbitrarily large before the FI is evaluated, appears to generate a result independent of $(3.5)$, which would be a most convenient simplification.

We cannot prove this conjecture with any mathematical rigour -- and it may be false -- but for the purposes of this paper it is not necessary. What is immediately clear is that the lowest order expansion of $(3.6)$ or $(3.7)$ will contribute nothing to the fourth order estimate of $\tilde K_{\mu\nu}(k)$, because a single $\delta/\delta\chi_{\lambda\mu}$ operating upon ${\rm tr}\,\Bigl(e\,^{\displaystyle i\int\!\!\sigma.\chi}\Bigr)_+\Big\vert_{\chi\rightarrow 0}$ produces the factor tr\,$[\sigma_{\lambda\mu}]=0$. $\tilde K_{\mu\nu}^{(4)}$ is therefore given precisely and exactly by the simple insertion of the $g_0^2$ order exponential term of $(3.3)$ under the integrals defining $\tilde K_{\mu\nu}^{(2)}(k)$ :
\beq{}&\displaystyle \tilde K_{\mu\nu}^{(4)}(k) = -g_0^4\int_0^{\infty}\!\!{ds\over s}\,e\,^{\displaystyle -ism^2}\,\int_0^s\!\!ds_1\int_0^s\!\!ds_2\,\int\!\!{d^4p\over(2\pi)^4}N\!\int\!d[u]\,e\,^{\displaystyle {i\over2}\int u(2h)^{-1}u} \\ & \times\displaystyle \biggl[{\partial\over\partial s_a}{\partial\over\partial s_b}{\delta\over\delta g_{\mu}(s_a)}{\delta\over\delta g_{\nu}(s_b)} + (k_{\mu}k_{\nu} - \delta_{\mu\nu}k^2)\biggr]\,e\,^{\displaystyle \!i\int_0^s\!\!\!u.(f+g)}\!\!\int_0^s\!\!d\bar s_1d\bar s_2\,u'(\bar s_1)\!\cdot\! u'(\bar s_2)D_c(u(\bar s_1) - u(\bar s_2))\nonumber \eeq
with $D_c$ again in the Feynman gauge.

All that is now needed is to insert a Fourier representation for $D_c(u(\bar s_1) - u(\bar s_2))$ :
\beq\displaystyle \int\!\!{d^4q\over(2\pi)^4}\,{e\,^{\displaystyle \!iq.(u(\bar s_1) - u(\bar s_2))}\over q^2-i\varepsilon} = i\!\int_0^{\infty}\!\!d\tau\int\!\!{d^4q\over(2\pi)^4}\,\,e\,^{\displaystyle \!-i\tau q^2 + iq.(u(\bar s_1) - u(\bar s_2))} \eeq
to replace the exp$\,[iq.(u(\bar s_1) - u(\bar s_2))]$ of $(3.9)$ by exp$\,[i\int_0^sj_{\mu}(s')u_{\mu}(s')]$ where $j_{\mu}(s') = q_{\mu}[\delta(s'- \bar s_1) - \delta(s'- \bar s_2)]$; to replace the $u'(\bar s_1)\!\cdot\! u'(\bar s_2)$ of $(3.8)$ by $(\partial/\partial \bar s_a)(\partial/\partial \bar s_b)\Bigl({1\over i}\Bigr)^2\bigl(\delta/\delta j_{\mu}(\bar s_a)\bigr)\bigl(\delta/\delta j_{\nu}(\bar s_b)\bigr)$; and to then perform the gaussian integrals over $u$ and $q$, before taking the limits $\bar s_{a,b}\rightarrow\bar s_{1,2}$. The result is a set of integrals over $\displaystyle\int_0^s\!\!ds_1\int_0^s\!\!ds_2\int_0^s\!\!d\bar s_1\int_0^s\!\!d\bar s_2\int_0^{\infty}\!\!d\tau$, to be followed by the final $\displaystyle\int_0^{\infty}\!\!{ds\over s}\,e\,^{\displaystyle -ism^2}$; and this computation should be vastly simpler than that of the original Jost--Luttinger calculation \cite{five}.

There are two, related situations where linkages operating on OE terms may be expected to give a zero contribution to all orders, and these arise in the Dominant Part ( DP )  Model of Section 4, and somewhat differently in the Extended DP Model of Section 5. In both of these cases we will be interested in a specific limit in which a $\Delta u_{\mu}$ variable approaches zero, and we will insist that these limits should be taken in a symmetric way, for both the odd function $\displaystyle\partial_{\lambda}\,D_{c}(\Delta u) = -{i\over2\pi^2}{\Delta u_{\lambda}\over[(\Delta u)^2+i\varepsilon]^2}$, and the even function $\displaystyle\partial_{\lambda}\partial_{\mu}\,D_{c}(\Delta u) = -{i\over2\pi^2}{1\over[(\Delta u)^2+i\varepsilon]^2}\,(\delta_{\mu\nu} - 4{\Delta u_{\mu}\Delta u_{\lambda}\over[(\Delta u)^2+i\varepsilon]})$. In such a symmetric limit, as $\Delta u\rightarrow 0$, we will employ :
\beq\displaystyle\partial_{\lambda}\,D_{c}(\Delta u)\Big\vert_{\Delta u\rightarrow 0} = 0\eeq
and :
\beq\displaystyle\partial_{\lambda}\partial_{\mu}\,D_{c}(\Delta u)\Big\vert_{\Delta u\rightarrow 0}\sim (\delta_{\mu\nu} - {4\over[(\Delta u)^2+i\varepsilon]}\cdot{1\over4}\,\delta_{\mu\nu}\,(\Delta u)^2) = 0\eeq
and so replace $(3.5)$ by unity, as well as the cross linkage operation between the OE terms of $\displaystyle{\delta^2 L\over\delta A_\mu\delta A_\nu}$ and exp$\,(L[A])$.
\bigskip
{\bf\section{The DP Model for a single closed fermion loop}}
\setcounter{equation}{0}
Return to the $\tilde K_{\mu\nu}^{(2)}(k)$ calculated in Section 2, written as :
\beq \displaystyle \tilde K_{\mu\nu}^{(2)}(k) = (k_{\mu}k_{\nu} - \delta_{\mu\nu}k^2)\Pi^{(2)}(k^2)\ ,\ \ \ \ \Pi^{(2)}(k^2) = \int_0^{\infty}\!\!{ds\over s}\,e\,^{\displaystyle -ism^2}\,\Pi^{(2)}(k^2,s)\eeq
where :
$$\displaystyle\Pi^{(2)}(k^2,s) = {g_0^2\over 2\pi^2}\int_0^1\!\!dy\,y(1-y)\,e\,^{\displaystyle -isk^2y(1-y)}$$
We now insert the missing exponential part of this integrand, exp$\,\displaystyle [i\,{g_0^2\over2}\int \!\!u'D_cu']$ coming from all the remaining terms of $\displaystyle e\,^{\displaystyle{\cal D}\!\!_A}\,{\delta^2 L\over\delta A\,\delta A}\Bigl|_{A=0}$ and retain only those parts which can contribute to subsequent UV divergences; these are the `` dominant '' parts, which define the DP Model, and in the following way.

As noted in Section 3, the choice of gauge is irrelevant, and we choose the simplest Feynman gauge, where the exponential factor $\displaystyle{i\over2}\int \!\!u'D_cu'$ becomes :
\beq\displaystyle -{g_0^2\over4\pi^2}\int_0^s\!\!ds_1\int_0^s\!\!ds_2\,u'(s_1)\!\cdot\!u'(s_2)\Bigl[(u(s_1) - u(s_2))^2+i\varepsilon\Bigr]^{-1}  \eeq
Since we are concerned in this and the following Section with a particular $\Delta u\rightarrow0$ limit, all subsequent OE terms will be discarded.

It is intuitively clear that the  most significant contributions will arise when $u(s_1)$ is close to $u(s_2)$, and we therefore expand $u(s_2)$ about $s_1$, writing :
$$u_{\mu}(s_2)\simeq u_{\mu}(s_1) - (s_1-s_2)u_{\mu}'(s_1) + \cdots$$
with $u(s_1)$ and $u'(s_1)$ considered as continuous functions. ( The continuity of $u(s')$ is clear from its definition, as the integral over the Fradkin 4 velocity $v(s')$; while the continuity of $u'(s') = v(s')$ follows from the physical expectation that the 4 velocity of a particle, real or virtual, must be treated as a continuous function of its proper time parameter.) All higher derivatives need not be continuous, and there is no obvious way of calculating them and their fluctuations; but they should not contribute to the leading divergent structures produced by the DP Model. This point is discussed and justified in detail in Appendix B.

To test the DP Model, in Appendix A we exhibit a completely independent, and simple perturbative example, whose log divergence is -- to within additive constants -- precisely the same as that calculated by the DP Model. But this point should be intuitively clear : because all $u$ fluctuations are controlled by the gaussian weighting of the Fradkin representation, all $u$ fluctuations must satisfy $u\le\sqrt s$; and because all UV divergences arise from small $s$, and therefore from small $s_1-s_2$, differences which scale as $s$, we may retain only the $(s_1-s_2)u'(s_1)$ part of the denominator of $(4.2)$, and replace the numerator $u'(s_2)$ by $u'(s_1)$. These simple replacements define the DP Model, and effectively permit the extraction of the leading divergence structure from under the FI of the $u$ fluctuations.

The DP Model thus replaces $(4.2)$ by the far simpler quantity :
$$\displaystyle-{g_0^2\over4\pi^2}\int_0^s\!\!ds_1\int_0^{s_1}\!\!ds'\,[u'(s_1)]^2\,\Bigl[s'^2u'^2(s_1)+i\varepsilon\Bigr]^{-1},\ \ \ \ s'=s_1-s_2,\ \ \ \ u'^2=u_{\mu}'u_{\mu}'$$ 
where it is understood that the effective UV cut off $\varepsilon$ is to be held fixed until the very last step of all calculations. If $[u'(s_1)]^2=0$, this integral vanishes; if not, with $\alpha_0 = g_0^2/4\pi$ it can be rewritten as :
$$\displaystyle-{\alpha_0\over\pi}\int_0^s\!\!ds_1\int_0^{s_1}\!\!ds'\,\Bigl[s'^2+i\varepsilon\bar\varepsilon(s_1)\Bigr]^{-1}$$
where we denote by $\bar\varepsilon(s_1)$ the sign of $[u'(s)]^2$. And since $\displaystyle\int_0^{s_1}\!\!ds'\,\Bigl[s'^2+i\varepsilon\bar\varepsilon(s_1)\Bigr]^{-1}$ can be rewritten as $\displaystyle\int_0^{s_1}\!\!ds'\,\Bigl[s'+i\varepsilon\bar\varepsilon(s_1)\Bigr]^{-2} = -{1\over s_1+i\varepsilon\bar\varepsilon} + {1\over i\varepsilon\bar\varepsilon}$, the first of the two needed integrations can be trivially performed.

We now insist that the second term immediately above, $\displaystyle{1\over i\varepsilon}\int_0^{s}\!\!ds_1{1\over\bar\varepsilon(s_1)} = {1\over i\varepsilon}\int_0^{s}\!\!ds_1\,\bar\varepsilon(s_1)$, must vanish, and for the following several reasons. (a) Intuitively, nothing in the formalism distinguishes between positive and negative values of $[u']^2$, and therefore the sum over the sign of all possible $[u']^2$ fluctuations must vanish. (b) If this integral did not vanish, there would appear in all orders of perturbation theory its coefficient $1/\varepsilon$, which corresponds to a quadratic UV divergence; and in all known examples of perturbative, gauge invariant, QED calculations, such quadratic UV divergences are absent. (c) One can point to related, if indirect, arguments, such as treating the FI of the Fradkin representation as a normalized probability function, and using it to calculate the value of the expected sign of the integral of $u'^2$, which also turns out to be zero. Henceforth, we shall assume this most reasonable property, that the integral over the sign of $u'^2$ vanishes.

With this understanding, our desired exponential factor is modeled by :
$$\displaystyle{\alpha_0\over\pi}\int_0^{s}\!\!ds_1\,\Bigl[s_1+i\varepsilon\bar\varepsilon(s_1)\Bigr]^{-1}$$
Since we expect $\bar\varepsilon$ to fluctuate, as the integral over $s_1$ proceeds, and since the knowledge of $\bar\varepsilon$ is only necessary near the lower limit of the $s$ integral, we can rewrite the integral as :
\beq\displaystyle{\alpha_0\over\pi}\int_{\varepsilon}^{s}\!{ds_1\over s_1} = {\alpha_0\over\pi}\ln\Bigl({s\over\varepsilon}\Bigr)\eeq
which quantity misses a possible $\displaystyle{\alpha_0\over\pi}\ln(i\bar\varepsilon(0))$ additional term, depending on the sign of $\bar\varepsilon(s_1\rightarrow0)$. For the moment, we suppress this factor, retaining the obvious $\ln\Bigl({s/\varepsilon}\Bigr)$ dependence.

The exponential of $(4.3)$ is our DP Model of all the radiative corrections ( of this Section~) to $\Pi^{(2)}(k^2)$ and yields for the sum of those contributions :
\beq \displaystyle \Pi^{(2)}(k^2) = 2{\alpha_0\over \pi}\int_0^1\!\!dy\,y(1-y)\!\int_{\varepsilon}^{\infty}\!{ds\over s}\Bigl({s\over\varepsilon}\Bigr)^{\alpha_0/\pi}\,e\,^{\displaystyle -is[m_0^2 + k^2y(1-y)]}\eeq
where we have cut off the lower limit of the $s$ integral at $\varepsilon$, and will later use the identification $\varepsilon\rightarrow1/\Lambda^2$. Although $(4.4)$ contains log divergent terms in every order, it is interesting to note that these additional radiative corrections conspire to remove the necessity of using an $\varepsilon$ as the lower limit o the $s$ integral if $\alpha_0>0$. In any finite order perturbation calculation, that lower limit $ \varepsilon$ cut off is absolutely necessary; but since we have summed over an infinite number of graphs, there has occurred a qualitative change in the $s$ integrand, such that it was not necessary to introduce this lower limit ( although the need for a cut off is simply transferred to large values of $s$ ). There are still divergences, in every order (~of an $\alpha_0$ expansion ), but this effect suggests the possibility that the omitted radiative corrections, arising from the linkage of all possible closed fermion loops to those loops under consideration above, may tend to suppress the divergences which remain in $(4.4)$. And this is indeed the case.

Restoring the neglected phase factor of $(4.3)$, we calculate $Z_3^{-1}=1+\Pi(0)$ by rotating the contour of the $s$ integration to run along the negative imaginary axis, and so obtain :
\beq \displaystyle \Pi(0) = {\alpha_0\over 3\pi}\int_{\varepsilon m_0^2}^{\infty}\!{d\tau\over\tau}\Bigl({\tau\over m_0^2\varepsilon}\Bigr)^{\alpha_0/ \pi}\,\exp\Bigl[\displaystyle {\alpha_0\over \pi}\ln(-i)-{\alpha_0\over \pi}\ln(i\bar\varepsilon(0))\Bigr]\eeq
and note that the reality of $Z_3$ fixes the choice $\bar\varepsilon(0)=-1$.

Changing variables, in $(4.5)$, to $\tau=\varepsilon m_0^2x$ produces :
$$\displaystyle \Pi(0) = {\alpha_0\over 3\pi}\int_1^{\infty}\!\!dx\,x^{\alpha_0/ \pi-1}\,e\,^{\displaystyle -\varepsilon m_0^2x}\simeq{\alpha_0\over 3\pi}\int_1^{1/\varepsilon m_0^2}\!\!dx\,x^{\alpha_0/ \pi-1}$$
or :
\be \displaystyle \Pi(0)\simeq {1\over 3}\Bigl[\Bigl({\Lambda^2\over m_0^2}\Bigr)^{\alpha_0\over\pi} - 1\Bigr]\simeq {1\over 3}\Bigl({\Lambda^2\over m_0^2}\Bigr)^{\alpha_0\over\pi}\ee
as long as $\alpha_0>0$ and $(\Lambda/m_0)\gg1$. $Z_3^{-1}$ can now be represented by $(4.6)$; or by an infinite sequence of log divergent terms :
\be\displaystyle Z_3^{-1} = 1 + {1\over 3}\sum_{n=1}^{\infty}{1\over n!}\Bigl({\alpha_0\over\pi}\Bigr)^n\ln^n\Bigl({\Lambda^2\over m_0^2}\Bigr)\ee
An alternate representation of $Z_3^{-1}$ is obtained by allowing the lower limit of the integral of $(4.5)$ to approach 0, which yields :
$$\displaystyle \Pi(0) = {\alpha_0\over 3\pi}\Bigl({\Lambda^2\over m_0^2}\Bigr)^{\alpha_0\over\pi}\int_0^{\infty}\!\!d\tau\,e\,^{\displaystyle -\tau}\,\tau^{\alpha_0/ \pi-1} = {1\over 3}\,\Bigl({\Lambda^2\over m_0^2}\Bigr)^{\alpha_0\over\pi}\,\Gamma(1 + {\alpha_0\over\pi})$$
and then :
\be\displaystyle Z_3^{-1} = 1 + {1\over 3}\,\Bigl({\Lambda^2\over m_0^2}\Bigr)^{\alpha_0\over\pi}\,\Gamma(1 + {\alpha_0\over\pi}) \ee
The gauge invariant part of the photon propagator, in this approximation, is given by :
$$\displaystyle\tilde D_c'(k)(\delta_{\mu\nu} - k_{\mu}k_{\nu}/k^2)$$
where :
\be\displaystyle\Bigl[\tilde D_c'(k)\Bigr]^{-1} = k^2\,[1+\Pi(k^2)]\ee 
and $\Pi(k^2)$ is the sum over those proper self energy terms ( which cannot be constructed from an iteration over lower order terms ) calculated in DP approximation. The simplest renormalization procedure in QED proceeds by adding and substracting $\Pi(0)$ in the inverse of the denominator of $(4.9)$, so that :
$$\displaystyle\tilde D_c'(k) = (k^2)^{-1}\Bigl[1+ \Pi(0) + \Pi(k^2) - \Pi(0) \Bigr]^{-1}$$
and the wave function renormalization constant $Z_3$ identified as the coefficient
 of the $k^2$ pole of $\tilde D_c'(k)$ as $k^2\rightarrow0$, which leads to the familiar identification of $Z_3$ in terms of $1+ \Pi(0)$. The renormalized propagator is : 
$$\displaystyle\tilde D_{c',R}(k) = Z_3^{-1}\,\tilde D_c'(k)$$
or :
\be\displaystyle\tilde D_{c',R}(k) = (k^2)^{-1}\Bigl[1+ Z_3(\Pi(k^2) - \Pi(0)) \Bigr]^{-1}\ee
where the combination $Z_3(\Pi(k^2) - \Pi(0))$ defines, in each sequential order, a finite contribution given in terms of the $\alpha$ constructed from the $Z_3$ contribution of that order, and $\alpha_0$ is chosen to have whatever ( large ) value is required so that the renormalized $\alpha=1/137$.

In our DP Model, the calculation of $\Pi(k^2) - \Pi(0)$ is again immediate, yielding :
\be\displaystyle\Pi(k^2) - \Pi(0) = -{1\over 3}\,\Bigl({\Lambda^2\over m_0^2}\Bigr)^{\alpha_0\over\pi}\,\Gamma(1 + {\alpha_0\over\pi})\biggl[1 - 6\int_0^1\!\!dy\,y(1-y)\Bigl(1 + {k^2\over m_0^2}\,y(1-y)\Bigr)\biggr]^{-{\alpha_0\over\pi}}\ee
and :
\be\displaystyle Z_3\Bigl[\Pi(k^2) - \Pi(0)\Bigr] = -\biggl[1 - 6\int_0^1\!\!dy\,y(1-y)\Bigl(1 + {k^2\over m^2}\,y(1-y)\Bigr)\biggr]^{-{\alpha\over\pi}}\ee
and where all the $\alpha_0$ dependence of $(4.11)$ has -- in any finite order -- been `` sequentially transformed '' into the $\alpha$ dependence of $(4.12)$.

In standard texts \cite{eight}, the finite order passage from $(4.11)$ to $(4.12)$ is emphasized, and correctly, as incorrect; but if the conventional, sequential renormalization procedure is to be maintained with the inclusion of an infinite number of divergent radiative corrections, the situation is quite different. Of course, the conventional procedure of renormalization has long been considered somewhat dubious, since it involves an effective perturbative expansion in powers of ( the arbitrarily large ) $\alpha_0$. But the procedure does give reasonable, renormalized expressions, such as its 27 $Mc$ contribution to the Lamb shift, and the detailed calculations \cite{eight} showing that the renormalized $\alpha^2$ corrections to this vacuum polarization tensor for large $(k/m)$ are proportional to $\ln(k/m)$ for both order $\alpha$ and $\alpha^2$, while the leading $\alpha^3$ term is proportional to the square of that logarithm.

But the fact that our DP result for large $(k/m)$ does not agree with the Jost-Luttinger perturbative result is irrelevant, for there is no reason to expect the DP model, which deals with the extraction of log divergences, to provide the correct limiting values of perturbative quantities, such as the coefficients of the leading, and finite, $\alpha^2\ln^2(k/m)$ dependence, although such a possibilty may be approximately realized. While our DP Model sums up all the possible perturbative divergences to the gauge invariant photon renormalization, it misses possibly important additive $k/m$ dependence to those divergent logs, for it is computing in the region of $k=0$, rather than large $k/m$. But the DP method does show, as in Section 5 for the $k=0$ computation of $Z_3^{-1}$, that it is sensible to consider this gauge invariant sector of QED as a finite QFT; and this, conceptually, is a new and most satisfying result.

Finally, to understand what it is that our DP Model achieves, we ask and answer the
following question. How can one understand the connection between the vanishing of
a propagator's denominator in this configuration space, functional formulation and
the appearance of log divergences in momentum space ? Simply by taking the Fourier
transform of $D_c(u(s_1) - u(s_2))$ and asking why does the corresponding 
$$\displaystyle \int\!\!{d^4k\over(2\pi)^4}\,{e\,^{\displaystyle \!ik.(u(s_1) - u(s_2))}\over k^2}$$

\noindent converge when combined with another momentum-space propagator? It is not the
insertion of a denominator factor proportional to $[(q-k)^2 + M^2]^{-1}$ coming from
another propagator, bosonic or fermionic, for that alone generates a log divergence.
 Rather, it is the $\exp[ik.(u(s_1) - u(s_2))]$ term -- which of course is eventually evaluated
in and by the $\int\!d[u]$ FI -- which provides enough oscillations and cancellations to
yield a finite result. But when $u(s_1) - u(s_2)$ vanishes, as we have suggested above, that
log divergence will appear and will enter all relevant parts of the computations of
that order. We find it far simpler to remain in functional configuration space,
where the cancellations of the OE terms are easily visible, then to convert to the
conventional Feynman graph analysis in momentum space. The divergent logs will
reappear, but the cancellations obvious in functional configuration space will there
require much tedious calculation to obtain. It is far more
 efficient to identify and extract in their functional source those log divergences,
than to perform the FIs, convert to momentum space, distribute those divergences in
the conventional parts of a Feynman graph, and then attempt, order by order, to
understand and remove them.
\bigskip
{\bf\section{The extended DP Model, and the finiteness of $Z_3$}}
\setcounter{equation}{0}
We return to (2.3) and the exact $K_{\mu\nu}$, from which follows the complete $Z_3^{-1}$   . The
first task to perform is to argue that the :
$$e\,^{\displaystyle{\cal D}\!\!_A}\,{\delta L\over\delta A_\mu(x)}{\delta L\over\delta A_\nu(y)}\,e\,^{\displaystyle L[A]} /\!< S >\Bigl|_{A=0}$$
terms cannot contribute to $Z_3^{-1}$, and for this we return to the Fradkin representation
for $L[A]$, here simplified by the neglect of all OE terms, as discussed in the
previous Section. Together with the Proof of Appendix B, this means that all OE
dependence of every $L[A]$ does not contribute to the radiative corrections calculated
in the DP Models.
\vskip0.2cm
Consider first the product $\displaystyle{\delta L\over\delta A_\mu(x)}{\delta L\over\delta A_\nu(y)}$, which contains under its separate integrals the terms relevant to this discussion:
\beq{}&\displaystyle \int\!\!d^4x'\delta(u(s))\!\int_0^s\!\!ds_1\,u_{\mu}'(s_1)\delta(x'-u(s_1)-x)\int\!\!d^4x''\delta(\bar u(\bar s))\!\int_0^{\bar s}\!\!d\bar s_1\,\bar u_{\nu}'(\bar s_1)\delta(x''-\bar u(\bar s_1)-y)\nonumber \\& \times\displaystyle\exp\left[-ig_0\!\int_0^s\!\!ds'\,u_{\alpha}'(s')\,A_{\alpha}( x' - u(s')) - ig_0\!\int_0^{\bar s}\!\!d\bar s'\,\bar u_{\beta}'(\bar s')\,A_{\beta}( x'' - \bar u(\bar s'))\right]\nonumber\eeq
Neglecting for the moment cross linkages to the $\exp(L[A])$ term, we then have :
\beq{}&\displaystyle e\,^{\displaystyle{\cal D}\!\!_A}\,{\delta L\over\delta A_\mu(x)}\,e\,^{\displaystyle\buildrel\leftrightarrow\over{\cal D}\!\!_A}\,e\,^{\displaystyle{\cal D}\!\!_A}{\delta L\over\delta A_\nu(y)}\Bigl|_{A=0}\,=\nonumber \\& \displaystyle\int\!\!d^4x'\delta(u(s))\int\!\!d^4x''\delta(\bar u(\bar s))\!\!\int_0^s\!\!ds_1\int_0^{\bar s}\!\!d\bar s_1\,u_{\mu}'(s_1)\!\,\bar u_{\nu}'(\bar s_1)\,\delta(x'-u(s_1)-x)\,\delta(x''-\bar u(\bar s_1)-y)\nonumber \\& \times\,
\displaystyle e\,^{\displaystyle i{g_0^2\over2}\int\!\!\!\int_0^s\!\!ds_1'ds_2'\,D_{c}(u(s_1') - u(s_2'))u'(s_1').u'(s_2')}\,e\,^{\displaystyle i{g_0^2\over2}\int\!\!\!\int_0^{\bar s}\!\!d\bar s_1'd\bar s_2'\,D_{c}(\bar u(\bar s_1') - \bar u(\bar s_2'))\bar u'(\bar s_1').\bar u'(\bar s_2')}\nonumber\\&
\times\,\displaystyle e\,^{\displaystyle ig_0^2\int_0^s\!\!d\hat s_1\!\int_0^{\bar s}\!\!d\hat s_2\,u'(\hat s_1).\bar u'(\hat s_2)\,D_{c}(x' - x'' + \bar u(\hat s_2) - u(\hat s_1))}\eeq
With the aid of the delta functions of (5.1), its last line may be rewritten as :
\be\displaystyle e\,^{\displaystyle ig_0^2\int_0^s\!\!d\hat s_1\!\int_0^{\bar s}\!\!d\hat s_2\,u'(\hat s_1).\bar u'(\hat s_2)\,D_{c}(x - y + \Delta(u, \bar u))} \ee
and the $\displaystyle\int\!\!d^4x'\!\!\int\!\!d^4x''$ performed so that (5.1) reduces to :
\beq{}& \displaystyle\delta(u(s))\delta(\bar u(\bar s))\!\!\int_0^s\!\!ds_1u_{\mu}'(s_1)\int_0^{\bar s}\!\!d\bar s_1\,\!\,\bar u_{\nu}'(\bar s_1)\,{\cal S}(\Delta u){\cal S}(\Delta \bar u)\nonumber \\& \times\,
\displaystyle e\,^{\displaystyle ig_0^2\int_0^s\!\!d\hat s_1\!\int_0^{\bar s}\!\!d\hat s_2\,u'(\hat s_1).\bar u'(\hat s_2)\,D_{c}(x - y + \bar u(\hat s_2) - u(\hat s_1))}\eeq
where $\Delta(u, \bar u)) = u(s_1) - u(\hat s_2)- \bar u(\bar s_1) + \bar u(\hat s_2)$, and ${\cal S}(\Delta u)$ and ${\cal S}(\Delta \bar u)$ refer to the self linkage exponentials of (5.1).

The renormalized charge in QED is conventionally defined by evaluating $\tilde K_{\mu\nu}(k)$ at $k=0$, corresponding to the definition of physically measured charge at large distances, specifically at distances large compared to the Compton wavelength $m^{-1}$; the radiative corrections occur at distances less than $m^{-1}$, and in the context of these calculations, this corresponds to evaluating $\tilde K_{\mu\nu}(x-y)$ at separations $x-y >> m^{-1}$. But all of the $x-y$ dependence of (5.3) lies in the $D_c$  corresponding to its cross-linked exponential factor, while the $u$ and $\bar u$ quantities, from their definitions scale as $\sqrt s\sim\sqrt {\bar s}\sim m^{-1}$.
It is then clear that in the limit of $x-y >> m^{-1}$, the $\Delta u$  and $\Delta \bar u$
dependence of $D_{c}(x - y + \Delta(u, \bar u))$ is effectively suppressed, and in this limit the factors $\int_0^{s}d\hat s_1\,u'(\hat s_1)$ and $\int_0^{\bar s}d\hat s_2\,\bar u'(\hat s_2) = 0$, so that this exponential factor completely disappears. In a similar way, after the $x'$ and $x''$ integrations have been performed, the entire contribution of (5.3) is itself proportional to similar factors, $\int_0^{s}ds_1\,u_{\mu}'(s_1)\int_0^{\bar s}d\bar s_1\,\bar u_{\nu}'(\bar s_1)$ which also vanish. 

Insertion of the cross linkages between (5.1) and the $A$ dependnce of $\exp(L[A])$ does not change this situation, for the limit of large $x-y$ in the cross linked terms can
be understood as the separate limits of $x\rightarrow\infty$ and $y\rightarrow-\infty$, so that these cross linkages also vanish. The computation then reduces to that of (5.3), as the
self-linkages of $\exp(L[A])$ are cancelled by the definition of $<S>$. The result is
that the entire quantity :
$$e\,^{\displaystyle{\cal D}\!\!_A}\,{\delta L\over\delta A_\mu(x)}{\delta L\over\delta A_\nu(y)}\,e\,^{\displaystyle L[A]} /\!< S >\Bigl|_{A=0}$$
does not contribute to $Z_3$.

The first term of (2.3) is the relevant quantity, and we first consider the
structure of the factor $e\,^{\displaystyle{\cal D}\!\!_A}\,e\,^{\displaystyle L[A]}$. It will be convenient to express the latter in terms of the functional cluster
expansion \cite{nine} : $$e\,^{\displaystyle{\cal D}\!\!_A}\,e\,^{\displaystyle L[A]} = \exp\Bigl[\sum_{n=1}^{\infty}{1\over n!}\,\,Q_n[A]\Bigr]$$
where $Q_n[A] = e\,^{\displaystyle{\cal D}\!\!_A}\Bigl(L[A]\Bigr)^n\Bigl|_{\,\rm connected}$, that is : $Q_1[A] = e\,^{\displaystyle{\cal D}\!\!_A}L[A] \equiv \bar L[A]$, $Q_2[A] = (e\,^{\displaystyle{\cal D}\!\!_A}L)(e\,^{\displaystyle\buildrel\leftrightarrow\over{\cal D}\!\!_A}-1)(e\,^{\displaystyle{\cal D}\!\!_A}L)$, etc. The reason for choosing this expansion is that the $Q_n$, $n>1$, can be estimated to
yield smaller values than does $Q_1$ (as well as being far more difficult to
calculate); their divergence structures are similar to that of $Q_1$, but, as
discussed in Appendix C, they play a smaller role in the overall calculation. $Q_1$  
alone is sufficient to remove all the perturbative log divergences of $Z_3^{-1}$.

Rather than repeat all the details of every equation in the next few paragraphs, we
shall simply present the added features that arise from the cross linkages between the $e\,^{\displaystyle{\cal D}\!\!_A}\,\displaystyle{\delta^2 L\over\delta A_\mu(x)\delta A_\nu(y)}$ of the previous Section and the $e\,^{\displaystyle{\cal D}\!\!_A}\,e\,^{\displaystyle L}\simeq e\,^{\displaystyle \bar L}$ of the present discussion. Note that all the self linkages of $e\,^{\displaystyle{\cal D}\!\!_A}\,e\,^{\displaystyle L[A]}/\!\!<S>\!\Bigl|_{\,0}$ simply disappear from the final result by virtue of the definition of $<S>$. We are therefore interested in :
\be\Bigl(\,e\,^{\displaystyle{\cal D}\!\!_A}\,\displaystyle{\delta^2 L\over\delta A_\mu(x)\delta A_\nu(y)}\,\Bigr)\,e\,^{\displaystyle\buildrel\leftrightarrow\over{\cal D}\!\!_A}\,e\,^{\displaystyle \bar L[A]}\Bigl|_{\,A\rightarrow 0}\ee 
which, continuing to use the variables $u_{\mu}(s')$ for the FI of $e\,^{\displaystyle{\cal D}\!\!_A}\,\displaystyle{\delta^2 L\over\delta A_\mu(x)\delta A_\nu(y)}$ corresponds to the insertion under the latter's FI the quantity :
\be \displaystyle\exp\left[\, -g_0\!\!\int_0^s\!\!ds'\,u_{\alpha}'(s')\int\!d^4w\,D_{c}(x' - u(s') - w)\,{\delta\over\delta A_{\alpha}(w)}\right]\,e\,^{\displaystyle \bar L[A]}\Bigl|_{\,A\rightarrow 0}\ee                                         
where we again hold to the Feynman gauge.

But the operation of (5.5) is just a translation operator, and regardless of what
it acts upon, has the effect of shifting the $A$ dependence of that function -- in
this case $\bar L[A]$ -- by the quantity :
\be \displaystyle\exp\left[\, ig_0^2\!\!\int_0^s\!\!ds'\,u_{\alpha}'(s')\int_0^t\!\!dt'\,v_{\alpha}'(t')\,D_{c}(x' - x'' - u(s') + v(t'))\,\right]\ee
appearing under the FI of $\bar L[A]$, with $x''$ and $v_{\alpha}(t')$ the variables of that
functional. In addition to the factor of (5.6), there appear under the $\bar L$ FI the
self linkages of amount :
$$\displaystyle \exp\left[\, i{g_0^2\over2}\int\!\!\!\int_0^t\!\!dt_1dt_2\,D_{c}(v(t_1) - v(t_2))v'(t_1).v'(t_2)\,\right]$$
which are independent of $x'$ and $x''$. In effect, what the cross linkages have
achieved is to insert $s$ dependence under and mixed with the $t$ integrals of $\bar L$; and
subsequent integration of that $t$ dependence generates an exponential $s$ dependence
which will have a damping effect on all the divergent expressions of Section 4. This
simple observation is at the heart of the mechanism for obtaining a finite charge
renormalization.

The self linkages of $\bar L$ may be read off from those calculated in Section 4, eq.(4.3) and its subsequent discussion :
$$\displaystyle \exp\left[\, i{g_0^2\over2}\int\!\!\!\int_0^t\!\!dt_1dt_2\,D_{c}(v(t_1) - v(t_2))v'(t_1).v'(t_2)\,\right] \rightarrow \Bigl({t\over i\varepsilon.\bar\varepsilon(0)}\Bigr)^p$$
The cross linkage term of (5.6) can be evaluated in a similar manner, with the
realization that only the continuous parts of the functions $u(s')$ and $v(t')$ are
relevant -- as noted in Appendix B -- and in essence they are very similar functions,
differing mainly in their physical place in the calculation. They represent the
continuous parts of the fluctuations defined by the same functionals; and it is
reasonable to ask when they can interact and combine directly with each other, for
when $u(s')\approx v(t')$ one has the beginning of an incipient divergence. A
modification of the previous DP model is now defined by expanding $v(t')\simeq v(s') + (t'-s')v'(s') + \cdots$, for the case when these functions are essentially the same: when $u(s')\simeq v(t')$ and $u'(s')\simeq v'(t')$, so that the denominator of the cross linkages becomes~:
\be \Bigl[\,((x'-x'') + (t'-s')v'(s'))^2 + i\epsilon\,\Bigr]\ee
or the same form with $v$ replaced by $u$. Since the functional integrations sum over all possible (continuous) forms, which fluctuations are defined in exactly the same way, there should be a strong possibility of such overlaps. The essence of this Extended DP Model is that $u(s')$ and $v(t')$ are treated in a completely symmetric manner; and this requirement of symmetry turns out to be a guarantee of simplicity of the forms that follow. This intuitive assumption defines the Extended DP Model, in which the difference of two, equivalent, continuous fluctuations of
identical functionals has the same possibility of overlap as in the DP Model. We emphasize that we cannot prove the validity of this assumption; but it is most certainly intuitive; and it forms  the basis of the cancellations of divergent logarithms that are about to occur.

However, even the EDP Model cannot guarantee the vanishing of (5.7), for another
difference, $x' - x''$, need not be small, and in most cases is not. What is the
consequence when $x'-x''$ is large ? Quite independently of the EDP Model, when 
$x'-x''>>m^{-1}$, the Compton wavelength of the charged fermion traveling about the loop,
that difference completely dominates $u(s') - v(t')$, because each of the latter quantities scale as $\sqrt s\sim\sqrt t\sim m^{-1}$, and hence the $u-v$ difference in that case is irrelevant. But then, as repeatedly emphasized, the $s'$ and $t'$ integrals of (5.6) vanish, $\displaystyle\int_0^s\!\!ds'\,u_{\alpha}'(s')\int_0^t\!\!dt'\,v_{\alpha}'(t') = 0$ and remove the entire cross-linkage term from consideration. When does this not happen ? Only when $x'-x''$ is restricted to values on the order of, or less than $u(s') - v(t')$. But we are interested in small differences, where $s'$ and $t'$ tend to the order of $\varepsilon$, and where subsequently $\varepsilon\rightarrow 0$. How can this be arranged ? 

We shall here assume that the only contribution to the cross linkage integral comes when $|x'-x''|\simeq\xi\sqrt\varepsilon$, where $\xi\sim O(1)$. Since this is an idealization, one must expect fluctuations about this condition, such that $\xi$ will turn out to be somewhat less than unity. For conceptual simplicity, choose the point $x'$ as the origin of the $x''$
coordinates, and consider a (Euclidean) 4-sphere of radius $\xi\sqrt\varepsilon$      
For any point within this sphere, $|x'-x''|$ effectively disappear from (5.7), and we
can apply the EDP Model; this means that the only non-zero values of the $x''$
integrals is given by :
\be \int\!\!d^4x''\rightarrow\int\!\!d\Omega_4\int_0^{\xi\sqrt\varepsilon}\!\!r^3dr = {\pi^2\over2}\,\xi^4\varepsilon^2\ee
Can (5.8) produce a non zero result ? Yes, because $\bar L$ is itself proportional to the
factors $\displaystyle -{1\over2}\,{\rm tr}[1]\int\!\!d^4x''\int_{\varepsilon}^{\infty}\!\!{dt\over t}\,e\,^{\displaystyle -itm^2}\,\,N\!\int\!d[v]\,\exp[\,{i\over2}\int v(2h)^{-1}v\,]\,\delta^{(4)}\Bigl( v(t)\Bigr)$ and, just as for the DP Model, the evaluation of $N\!\int\!d[v]$ yields : $\displaystyle N\!\int\!d[v]\,\exp[\,{i\over2}\int v(2h)^{-1}v\,]\,\delta^{(4)}\Bigl( v(t)\Bigr) = -i{\pi^2\over t^2}{1\over (2\pi]^4}$, so that the entire set of exponential integrals reduces to :
\be \displaystyle T\Bigl({s\over\varepsilon}\Bigr) = i\Bigl({\xi\over2}\Bigr)^4\varepsilon^2\,\int_{\varepsilon}^{\infty}\!\!{dt\over t^3}\,e\,^{\displaystyle -itm_0^2}\,e\,^{\displaystyle 3i\pi p/2}\,\Bigl({s\over\varepsilon}\Bigr)^p\Bigl({t\over\varepsilon}\Bigr)^{2p}\ee
In (5.9), one factor of $\displaystyle e\,^{\displaystyle i\pi p/2}\,\Bigl({t\over\varepsilon}\Bigr)^p$ arises from the self linkages of $\bar L$, while the cross linkages generate the remaining factors. We have maintained strict $s$, $t$ symmetry by writing :
\beq&\displaystyle \int_0^s\!\!ds'\int_0^{t}\!\!dt'\,{u'(s')\!\cdot\!v'(t')\over (u(s')-v(t'))^2+i\varepsilon} = \displaystyle \int_0^s\!\!ds'\int_0^{t}\!\!dt'\left[{\theta(s'-t')\over (s'-t'+i\varepsilon\bar\varepsilon(s'))^2}+{\theta(t'-s')\over (t'-s'+i\varepsilon\bar\varepsilon(t'))^2}\right]\nonumber \\&\displaystyle=\int_0^{s}\!\!ds'\left[{1\over i\varepsilon\bar\varepsilon(s')}-{1\over s'+i\varepsilon\bar\varepsilon(s')}\right] + \int_0^{t}\!\!dt'\left[{1\over i\varepsilon\bar\varepsilon(t')}-{1\over t'+i\varepsilon\bar\varepsilon(t')}\right] \nonumber\eeq
Again, the integrals over the $\bar\varepsilon$ factors vanish, and the result is simply : 
$$-\ln\Bigl({s\over i\varepsilon\bar\varepsilon(0)}\Bigr) -\ln\Bigl({t\over i\varepsilon\bar\varepsilon(0)}\Bigr)$$
which is properly symmetric in $s$ and $t$. Were that symmetry not preserved, the results would lead to far more complicated forms, requiring detailed numerical integrations in order to verify the expectations of a finite $Z_3^{-1}$ and an $\alpha$ close to $1/137$. In contrast, the symmetric Extended DP Model adopted here leads to results  
obtainable in closed form, and to the immediate verification of our expectations.

With the inclusion of these cross linkages, the DP integral representation for $Z_3^{-1}$ is changed to :
\be\displaystyle Z_3^{-1} = 1 + \Bigl({p\over3}\Bigr)\,e\,^{\displaystyle i\pi p/2} \,\int_{\varepsilon}^{\infty}\!\!{ds\over s}\,e\,^{\displaystyle -ism_0^2}\,\Bigl({s\over\varepsilon}\Bigr)^p\,e\,^{\displaystyle T(s/\varepsilon)}\ee
Note that we are keeping to the conventional perturbative form (although in
configuration, rather than momentum space) of cutting off all proper time integrals
with a lower limit of $\varepsilon$, which will shortly be set equal to zero. Without the
cross linkage factors that produce $T$, (5.10) has the same divergences as does (4.5);
but an entirely new situation now arises with the insertion of (5.9) into (5.10).
Proper time contours need not be rotated; all that is needed is the simple change of
variables: $s/\varepsilon = x$, $t/\varepsilon = y$, so that (5.10) may be rewritten as :
\be\displaystyle Z_3^{-1} = 1 + \Bigl({p\over3}\Bigr)\,e\,^{\displaystyle i\pi p/2} \,\int_{1}^{\infty}\!\!dx\,x^{p-1}\,e\,^{\displaystyle -ixm_0^2\varepsilon}\,e\,^{\displaystyle T(x)}\ee
where :
$$ \displaystyle T(x) = i\Bigl({\xi\over2}\Bigr)^4\,e\,^{\displaystyle 3i\pi p/2}\,x^p\,\int_{1}^{\infty}\!\!{dy\over y^3}\,e\,^{\displaystyle -i\varepsilon ym_0^2}\,y^{2p} $$                                  
and one notes that, for convergence as $y\rightarrow\infty$, one must have $p < 1$. For $p > 0$, $T(x)$ can act as a damping or oscillating factor that provides convergence for the $x$ integral as $x\rightarrow\infty$; and for both the $x$ and $y$ integrals, we may now safely let $\varepsilon\rightarrow 0$, and the divergences have disappeared. 

As this program is carried out, one notes the independence of $Z_3$ on $m_0$, or on $m$. This property is not at all clear from perturbation theory, where sequential renormalization of mass and charge, along with simultaneous changes in $Z_{1,2}$        
must appear. In fact, $Z_3$, and therefore $\alpha$, are independent of the charged
particle's mass, in agreement with the experimental fact that all charged fermions
obeying QED (but not simultaneously QCD) have the same electric charge.
 
What remains is to insure that $Z_3^{-1}$ is real, and hence the condition Im $Z_3^{-1} = 0$ specifies a relation that $p = \alpha_0/\pi$ must satisfy. In principle, this is
true; and if that condition leads to a single allowed value of $\alpha_0$, and hence
of $\alpha$, one will have solved an old and deep question in Physics. In our calculation, however, there appears the parameter $\xi$, as a
measure of the difficulty and uncertainty of extracting the divergent character of
(5.7). And with the neglect of higher terms of the cluster expansion, there is no
guarantee that a single value of $\alpha_0$ will emerge. Rather, with the cluster
approximation already made, and with those approximations we are about to make, we
are gratified to find a range of $p$ values, $0<p<1$, within which we can choose $\alpha_0$ such that $\alpha\sim1/137$.  

For $p<1$, the integral defining $T(x)$ is readily obtained, and the expression for $Z_3^{-1}$ now reads :
\be\displaystyle Z_3^{-1} = 1 + \Bigl({p\over3}\Bigr)\,e\,^{\displaystyle i\pi p/2} \,\int_{1}^{\infty}\!\!dx\,x^{p-1}\,e\,^{\displaystyle -qx^p}\ee
where the $q$ parameter is given by : 
$$q = -i\Bigl({\xi\over2}\Bigr)^4\,{e\,^{\displaystyle 3i\pi p/2}\over 2(1-p)}$$
Note that convergence of the $x$ integral is obtained for Re$(q)>0$, which corresponds to $p<2/3$, and that a necessary but not sufficient condition for the removal of Im$Z_3$ is that $p>1/3$.

The $x$ integral is completely trivial because its integrand is a perfect differential, and one finally obtains :
\be\displaystyle Z_3^{-1} = 1 + {2\over 3}\,(1-p)\,\Bigl({\xi\over2}\Bigr)^{-4}\,i\,e\,^{\displaystyle -i\pi p} \,e\,^{\displaystyle -q}\ee
Since we expect $\xi$ to be somewhat less than 1, the ration $(\xi/2)^4$ should be very small, and a good approximation to (5.13) is then :
\be\displaystyle Z_3^{-1} = 1 + {2\over 3}\,(1-p)\,\Bigl({\xi\over2}\Bigr)^{-4}\,\Bigl(\sin(\pi p) + i\cos(\pi p)\Bigr)\ee
and it is clear that $p=1/2$ insures that the $Z_3^{-1}$ of (5.14) is real, while the choice $\xi\simeq 0.397$ leads to an $\alpha = \pi p Z_3$ of approximately $1/137$. From this solution, one calculates $(\xi/2)^4\simeq 0.00155$, so that any correction to these parameters obtained by the use of (5.13) rather than (5.14) cannot differ from the above values of $p$ and $\xi$ by more than a few parts per thousand.

\bigskip
{\bf\section{Summary}}
\setcounter{equation}{0}
The thrust of this paper has been to argue, by summing the `` naturally
divergent '' terms of all relevant radiative corrections,  that charge renormalization
in QED is finite. We do not claim to have given a mathematically rigorous proof of
that statement, but rather an intuitive statement, based on the functional structure
of QED. We have argued that our extraction of logarithmically divergent terms
corresponds to those found in lower order radiative corrections using Feynman graph
techniques; and we believe that in momentum space, graphical techniques become
impossible, and therefore irrelevant, in any attempt to include all, or almost all,
radiative corrections of arbitrarily high order.

The functional techniques we use are based upon a convenient rearrangement of the
Schwinger/Symanzik functional solution for the generating functional of QED,
together with a slight rearrangement of Fradkin's most useful functional
representation for Green's functions $G_c [A]$ and, in particular, for $L[A]$, the log
of the so called `` fermion determinant~''. $L[A]$ contains the basic, gauge invariance
of the photon propagator, and that structure is here realized by means of most
convenient linkage operations. 

If a perturbative expansion is desired, this functional approach will exactly
reproduce the conventional Feynman graphs, but it has the great advantage of
working in configuration rather than momentum space, and one can take advantage of
cancellations which occur there before any computation is required, but which are
achieved in momentum space only after painful and tedious manipulations \cite{one}. And, it should be noted, that frequently, as is the case for
the lowest order radiative corrections to the photon propagator, singularities of
the Feynman integrals can mask the symmetry structure of the theory, an unpleasant
attribute of that method of calculation which is quite absent from functional
methods built around proper time representations\footnote{Functional representations provide a simple realization of the `` Last Rule '', often
stated and rarely understood, for writing the total of all Feynman graphs of a given
order: `` Sum over all topologically distinct graphs ''.}.

Within this formalism, and the intuition we have used to obtain the results
described above, we have derived a pair of integrals in (5.11), with $\varepsilon$  set equal to zero, which should express the finite character of charge renormalization. And we
have illustrated a possible solution of those equations with a simplified
model which generates a specific value of $\alpha_0$ that leads to an $\alpha\sim1/137$.
What this model solution does not do is to obtain a single, precise  and necessary value of
$\alpha_0$, and so determine $\alpha$.  As noted in the text, because of the
imprecision in extracting the divergences from the cross linked, closed fermion loop functionals, as expressed by the parameter $\xi$, this goal may be elusive; and we will have to be content with choosing an $\alpha_0$ which does reproduce the experimental $\alpha$. 

Previous attempts have been made some decades ago \cite{four} to simultaneously display
a cancellation of divergences, which might lead to a value of $\alpha$ close to its
renormalized value. These were noble efforts, especially in the context of Feynman
graphs; and one can now see why they were unsuccessful, for the crucial aspect of
including an infinite number of closed fermion loops, each containing  all possible
photonic "dressing" in a manifestly gauge invariant way, could not be done.  It is
gauge invariance, built into the Fradkin representation for $L[A]$, along with the
use of proper time techniques which preserve that invariance, and allows one to
identify and extract divergences. 

It should be noted also that the assumptions made in these papers, which center
about a single log divergence for $Z_3^{-1}$ persisting in higher perturbative orders,
together with the special choice of a zero bare fermion mass, are quite different
from what we observe. Our $Z_3^{-1}$ is finite and independent of mass. The use of a
finite number of terms in a Bethe--Salpeter kernel (cf ref[10], Section 1) misses an
infinite class of Feynman graphs, for a new class of graphs appears in each higher
order; and in each missing class there are an infinite number of Feynman graphs for
higher and higher orders. In contrast, we make no assumptions about the nature of
perturbative divergences in $Z_3^{-1}$;  we include them as we see them, and we are
able to see and include them because we use a formulation where their appearance is
obvious, and where strict gauge invariance helps us to extract them, and sum them
to all orders, literally.

There are other, relevant questions which immediately come to mind, such as why one should expect the approximate value of $\alpha$ to be close to $1/137$, when no mention has been made of electroweak symmetry, in which radiative corrections involving the weak interactions should play a role. Were we calculating a process at significant momentum transfers, or at energies on the order of the $W$ mass, for example, then our simple restriction to electron and photon QED would be insufficient. But the $Z_3^{-1}$ computation involves only radiative corrections evaluated at $k=0$, a most non--relativistic limit; and even though UV divergences may appear in every perturbative term, when summed, the result is independent of mass, as it becomes finite. Because the weak interactions are far weaker than electromagnetic interactions, we do expect corrections to our approximate calculations of $\alpha$ to be relatively small. 

It might also be noted that, from general principles enunciated long ago by Schwinger, one expects the renormalized $\alpha$ to be smaller than the unrenormalized $\alpha_0$; and since our result for the latter is appproximately $\pi/2$, it is a happy circumstance that the choice of $\xi\sim.4$ -- which, as expected, is on the order of but somewhat less than $1$ -- does produce an $\alpha$ close to its measured value.

Another immediate question concerns the possibility of employing the present methods to evaluate $Z_2$; can the analysis of the previous sections suggest, for any choice of relativistic gauge, that $Z_2^{-1}$ is finite ? Unfortunately, this does not seem to be possible, and the reason is that the conditions stated in section 5 for the validity of the EDP Model are no longer satisfied. Instead of the properties $u(s) = u(0) = v(t) = v(0)$ of the photon calculation, which allowed one to expect the similarity of the functions $u(s')$ and $v(t')$, for the electron propagator one has $u(0) = v(t) = v(0) = 0$, but $u(s) = -z = -(x-y)$, the variable conjugate to the electron's momentum. And as one goes to the mass--shell in $p^2$, one expects $z^2$ to become large, so that there will be a considerable difference in the (continuous) variations of $u(s')$ and $v(t')$.

For this reason, the intuition of the EDP Model disappears, and there remains no obvious mechanism to cancel the divergencies, as was done by the function $T(x)$ for $Z_3^{-1}$. On the basis of this argument, we see no alternative to the conclusion expressed long ago by K\"allen \cite{thirteen}, who conjectured that at least one of the renormalization constants of QED is divergent. From our perspective, it is the gauge dependent, unmeasurable $Z_2^{-1}$ which diverges, while the gauge independent $Z_3^{-1}$ is finite.

Perhaps the best way to end this paper is by emphasizing that while the functional
tools used are surely powerful, and appropriate for this problem, the
identification and extraction of divergences we have used has been intuitive. Nevertheless,
the obvious advantages of this functional approach to the calculation of radiative
corrections seems clear, and, we believe, deserves strong emphasis.
\vskip1truecm {\bf Acknowledgments} 
\vskip0.3truecm 
We owe thanks to many colleagues and friends for their patience and constructive
criticism; and in particular, to Walter Becker, Thierry Grandou, Alex Grossmann and Mark Rostollan.

This publication was made possible through the support of a grant from the John Templeton Foundation. The opinions expressed in this publication are those of the authors and do not necessarily reflect the views of the John Templeton Foundations.

\vskip1.5cm
\def\theequation{A.\arabic{equation}}
{\bf\section*{Appendix A. Equivalence Example of the DP Model}}
\setcounter{equation}{0}
\appendix
To demonstrate the accuracy of the DP Model in a quite different setting, consider the calculation of the simplest fermion self energy graph, where -- to make connection with the boson nature of the $L[A]$ calculations -- we simplify by passing to a bosonized version of relevant fermion propagator at the same stage of each calculation. This example shows that the log divergences of both calculations are exactly the same.

The $g^2$ order contribution to the fermion propagator is given, in functional notation, by~:
\be\displaystyle-{i\over2}\!\int\! {\delta\over\delta A_\mu}D_{c}^{\mu\nu}{\delta\over\delta A_\nu}\,G_c(x,y\,| A)\Big\vert_{A\rightarrow 0}\ee
\be\displaystyle=(-i)(ig)^2\!\int\! D_{c}(u-w)\,G_c(x,u\,| A)\gamma_{\mu}\,G_c(u,w\,| A)\gamma_{\mu}\,G_c(w,y\,| A)\Big\vert_{A\rightarrow 0}\ee
where we again adopt the simplest Feynman gauge $D_{c,\mu\nu}=\delta_{\mu\nu}D_c$

For the Feynman graph calculation, we replace both of the external $G_c$ by their free field limit $S_c$, and, to have a clear correspondence with the DP calculation, imagine that the central $G_c$ is replaced by a bosonized version, $<\!u\,|\bigl[ m^2 + (\partial - igA)^2\bigr]^{-1}|\,w\!>$, whose free particle limit is simply $<\!u\,|\bigl[ m^2 + \partial^2\bigr]^{-1}|\,w\!>$. Taking Fourier transforms, one finds the corresponding contribution to (A.2) :
$$\displaystyle{ig^2\over(2\pi)^4}\,\tilde S_c(p)\,\gamma_{\mu} \int\!\!d^4k\,{1\over k^2-i\varepsilon}\,{1\over (p-k)^2+m^2}\,\gamma_{\mu}\,\tilde S_c(p)$$
or, with $\displaystyle\tilde S_c(p) = {1\over m + i\slash \!\!\!\!p}\ $, $\displaystyle\sum_{\mu}\gamma_{\mu}\gamma_{\mu} = 4$ :
\be\displaystyle-{ig^2\over 4\pi^4}\left({1\over p^2}\right)\int\!\!d^4k\,{1\over k^2-i\varepsilon}\,{1\over (p-k)^2-i\varepsilon}\ee
where, for simple comparison with the DP Model, we choose $m=0$.

Using the simplest regularization possible :
$$\displaystyle{1\over k^2}\longrightarrow{1\over k^2} - {1\over k^2+\Lambda^2} = \int_0^{\Lambda^2}\!\!\!dl\,{2l\over (k^2 + l^2)^2}\,,\ \ \ \ \Lambda\rightarrow+\infty$$
and the Feynman combinatoric $\displaystyle{1\over a^2b}=2\int_0^1\!\!dx(1-x)[a+x(b-a)]^{-3}$, the integral of (A.3) is easily shown to be : $\displaystyle i\pi^2\int_0^1\!\!dx\ln\Bigl({\Lambda^2\over xp^2}\Bigr)$.

The result of this computation is then :
\be\displaystyle {g^2\over4\pi^2}{1\over p^2}\int_0^1\!\!dx\ln\Bigl({\Lambda^2\over xp^2}\Bigr)\ee
\indent We now calculate the equivalent quantity using the DP Model of the fermion propagator.

We recall the expression for the electron propagator \cite{three} :
$$\matrix{{}&\displaystyle G_c(x,y|A)=i\int_0^{\infty}\!\!ds\,e\,^{\displaystyle -ism^2}\,\,N\!\int\!d[u]\,e\,^{\displaystyle {i\over2}\int u(2h)^{-1}u}\,\displaystyle\biggl( m - \gamma\!\cdot\!{\delta\over\delta u'(s)}\biggr)\cr& \times\, \delta^{(4)}\Bigl( u(s) + x-y \Bigr)\,e\,^{\displaystyle -ig\int_0^s\!\!ds'\,u_{\mu}'(s')\,A_{\mu}( y - u(s'))}\,\biggl( e\,^{\displaystyle g\int_0^s\!\!ds'\,\sigma_{\mu\nu}\,F_{\mu\nu}( y - u(s'))}\biggr)_+}$$
We suppress all spinorial terms, set $m=0$ and $z=x-y$ . The $g^2$ order contribution to the fermion propagator is then :
\beq{}&\displaystyle i\int_0^{\infty}\!\!ds\,N\,\int\!d[u]\,e\,^{\displaystyle {i\over2}\int u(2h)^{-1}u}\,\delta^{(4)}\Bigl( u(s) + z \Bigr)\nonumber \\&\times\displaystyle {ig^2\over2}{i\over 4\pi^2}\int_0^s\!\!ds_1\int_0^s\!\!ds_2\,u'(s_1)\!\cdot\!u'(s_2)\Bigl[(u(s_1) - u(s_2))^2+i\varepsilon\Bigr]^{-1} \eeq
Using the notation and operations as defined in the text, the DP Model replaces the second line of (A.5) by :
\be\displaystyle {g^2\over 4\pi^2}\int_0^s\!\!ds_1\int_0^{s_1}\!\!ds'\,{\partial\over\partial s'}{1\over s'+i\varepsilon\bar\varepsilon} = \displaystyle {g^2\over 4\pi^2}\int_0^s\!\!ds_1\left[{1\over s_1+i\varepsilon\bar\varepsilon}-{1\over i\varepsilon\bar\varepsilon}\right]\ee
We again argue that $\displaystyle\int_0^s\!\!ds_1\,\bar\varepsilon = 0$, and replace the integral of (A.6) by :
\be\displaystyle\int_{i\varepsilon\bar\varepsilon(0)}^s\!\!{ds_1\over s_1} = \ln\left({s\over i\varepsilon\bar\varepsilon(0)}\right)\ee
The first line of (A.5) is then immediately given by : $\displaystyle i\!\int_0^{\infty}\!\!ds\int\!\!{d^4p\over(2\pi)^4}\,\,e\,^{\displaystyle ip.z }\,e\,^{\displaystyle -isp^2}$, and in combination with (A.7), its Fourier transform then yields :
\be\displaystyle {g^2\over 4\pi^2}\,\,i\!\int_0^{\infty}\!\!ds\,e\,^{\displaystyle -isp^2}\,\ln\left({s\over i\varepsilon\bar\varepsilon(0)}\right) \ee
Rotating the $s$ integration contour so that $s\rightarrow -i\tau$, and with $\varepsilon = \Lambda^{-2}$, $\bar\varepsilon(0)=-1$, $x=\tau p^2$, this becomes :
\be\displaystyle {g^2\over 4\pi^2}\,{1\over p^2}\,\,\int_0^{\infty}\!\!ds\,e\,^{\displaystyle -x}\,\ln\left({\Lambda^{2}x\over p^2}\right) \ee
and one sees that the log divergent terms of (A.9) and (A.4) are exactly the same.
\vskip1.5cm
\def\theequation{B.\arabic{equation}}
{\bf\section*{Appendix B}}
\setcounter{equation}{0}
\appendix
We can attempt a heuristic justification of the DP Model in the following way. Because the Fradkin $v(s')$, $0<s'<s$, refers to the four velocity of a virtual particle ( e.g., the fermion moving in a closed loop of the radiative correction corresponding to the simplest vacuum bubble ), we demand that $v(s')$, and therefore $u'(s')$, be a continuous function of its proper time, $s'$. This is a physical restriction on the class of functions $u(s')$ allowed; whether the particle is real or virtual, its four velocity should be and will be assumed to be continuous.

From its definition, $u(s') = \int_0^{s'}\!\!ds''v(s'')$, $u(s')$ itself must be a continuous function of $s'$; and hence we have restricted consideration to the class of functions $u(s')$ which are continuous and have a continuous first derivative. But no statement can be made about higher derivatives, which must be expected to be discontinuous. One way to describe such functions is to imagine that they begin life as continuous, with continuous second and higher derivatives proportional to finite constant $p_n$. Then, as the functional integration proceeds, imagine that certain of these parameters $p_n$ are changed, in a random way, to have values which approach $\pm\infty$; and this corresponds to the introduction of discontinuous  higher order derivatives. What will then be the effect of such fluctuations on the relevant functions :
$$\displaystyle -{g_0^2\over2 (2\pi)^4}\int_0^s\!\!ds_1\int_0^s\!\!ds_2\,u'(s_1)\!\cdot\!u'(s_2)\Bigl[(u(s_1) - u(s_2))^2+i\varepsilon\Bigr]^{-1}$$
of our functional integrand ?

Consider first the denominator $\Bigl[(u(s_1) - u(s_2))^2+i\varepsilon\Bigr]^{-1}$. Before the $p_n$ fluctuations are allowed to destroy the continuity of the second and higher derivatives, the main contribution of this denominator will, clearly, come from the expansion we have used, replacing $(u(s_1) - u(s_2))$ by $u'(s_1)(s_1-s_2)$, and neglecting terms such as $(s_1-s_2)^2u''(s_1)/2$. If the $p_n$ are then allowed to fluctuate such that any higher derivatives approach $\pm\infty$, the denominator will then become infinite, and so conveniently removes itself from the FI.

Now consider the $u'(s_1)\cdot u'(s_2)$ numerator term, which we have replaced by $u'^2(s_1)$. Again imagine that all derivatives begin  life as continuous, and consider the first correction to this approximation, $(s_2-s_1)u'(s_1)\cdot u''(s_1)$. If $u''(s_1)$ is finite, in the neighborhood $s_2\simeq s_1$, where the denominator is about to vanish, then this term generates a relatively small correction to our approximation; but if and when $u''(s_1)$ becomes infinite, there will be no contribution to the FI because the denominator fuctuations will involve $u''(s_1)^2$, which is more divergent than the numerator $u'(s_1)\cdot u''(s_1)$, and hence, both numerator and denominator contributions from such a discontinuous second derivative are removed from the FI.

This argument can be repeated for every higher order derivative, and for the sum of all such higher order derivatives; and the result is the DP Model defined in the text. This argument is heuristic, a physicist's argument; it is intuitive, rather than mathematically rigorous. To justify this intuition, we point to the example of Appendix A, wherein the log divergence calculated from the DP Model yields exactly the same result as that obtained from the corresponding Feynman graph integral. In this very real sense, a divergent Feynman graph in momentum space, calculated from continuous, if overlapping, momentum space integrands, may be thought of as equivalent to the `` continuous '' elements of a functional integrand in the DP Model.
\vskip1.5cm
\def\theequation{C.\arabic{equation}}
{\bf\section*{Appendix C}}
\setcounter{equation}{0}
\appendix
There are two distinct reasons why the connected $Q_n$, $n>1$, may be expected to generate a contribution to $T(x)$ which is considerably less than that of $Q_1$. The first is because of the reduced probability of finding overlaps of three or more coordinate systems whose origins must lie within a distance ( in Euclidean 4 dimensions ) of $\xi\sqrt\varepsilon$ from each other, one of those origins being that of the $x'$ of $\displaystyle{\delta^2 \bar L\over\delta A_\mu(x)\delta A_\nu(y)}$. The probability of such overlaps can be thought of as proportional to the overlapping volume divided by the total volumes of all three or more spheres each of radius $\xi\sqrt\varepsilon$, and that ratio is a small number, becoming smaller as $n$ increases. In mathematical terms, a non zero intersection of the supports of three or more independent distributions such as $D_cD_c\cdots D_c$ is much less than the corresponding quantity for the case of two such independent distributions.

The second reason is more closely tied to the computations of Section 5, and has as its origin the nature of the `` connectedness '' requirement, which can be stated in the following way. Connected linkages require that there shall be at least one linkage between the connected parties, e.g., for $Q_2$ :
\be \bar L\Bigl( e\,^{\displaystyle\buildrel\leftrightarrow\over{\cal D}\!\!_A} - 1 \Bigr)\bar L\,\Big\vert_{A\rightarrow 0}\ee
with the result that the linkages between the factors of $\displaystyle\exp\left[-ig_0\!\int_0^t\!\!dt'\,v'(t')\!\cdot\!A( x'' - v(t')) \right]$ from one $\bar L$, and the factor $\displaystyle\exp\left[-ig_0\!\int_0^r\!\!dr'\,w'(r')\!\cdot\!A( x''' - w(r')) \right]$ from the other $\bar L$, will appear in the form :
$$\displaystyle\exp\left[\,i g_0^2\int_0^t\!\!dt'\!\int_0^r\!\!dr'\,w'(r')\!\cdot\!v'(t')\,D_{c}(x'' - x''' + w(r') - v(t'))\,\right] - 1$$
which can be rewritten as :
$$\int_0^1\!\!d\lambda\,{\partial\over\partial\lambda}\exp\left[\,i\lambda g_0^2\int\!\!\!\int w'\!\cdot\!v'\,D_{c}\,\right]$$
or as :
\be \left[\,ig_0^2\int\!\!\!\int w'\!\cdot\!v'\,D_{c}\,\right]\,\int_0^1\!\!d\lambda\,\exp\left[\,i\lambda g_0^2\int\!\!\!\int w'\!\cdot\!v'\,D_{c}\,\right] \ee
Assume that the overlaps have taken place and consider the multiplicative term of $(C.2)$. It does not invovlve $y_{1,2}$ dependence raised to a power, but rather the logarithm of that dependence, which under the same variable changes as used in Section 5, will convert to $x$ and $u_{1,2}$ inside logs. From our over simplified model of Section 5, one sees that the significant $x$ value of the final $Z_3^{-1}$ integral is just barely larger than 1, and hence the $\ln(x)$ terms of this log dependence will not contribute significantly. The remaining $u_{1,2}$ factors must be evaluated within the $\int du_{1,2}$ of each $\bar L$. For $1<u_{1,2}<x$, we get another $\ln(x)$, but for $0<u_{1,2}<1$, there will be non zero contributions, integrable quantities of $O(1)$, relative to the results of the $\int du_{1,2}$ integrals without such terms.

However, these log terms multiply an integral which is essentially an average over values of $0<\lambda<1$, over the forms $(q)^{\lambda p}$, where $q$ takes on the different values which can be read off from the text, e.g. $\displaystyle(y_1)^{\lambda p}\Bigl(\,{y_1-i\over y_1-y_2-i}\,\Bigr)^{\lambda p}$. Since $p$ has been here replaced by $\lambda p$, for small $\lambda$ the contributions essentially disappear; and only when $\lambda\sim 1$ is there a significant value to the integral. This represents an effective decrease of the effectiveness of the coupling between the two $\bar L$s; and when multiplied by the small log terms discussed above, in addition to the small overlap factors, it seems clear that the connected terms cannot significantly add to $T(x)$, and therefore cannot significantly change the result of the $Q_1$ computation. Without a detailed and rigourous analysis, one cannot be absolutely sure; but this is our (intuitive) belief.

\end{document}